\documentclass[letter, 12pt]{article}

\usepackage[margin=1in]{geometry}
\usepackage[numbers,sort&compress]{natbib} % Citations
\usepackage{authblk} % Author info
\usepackage{color}
\usepackage{amsmath}
\usepackage{amssymb} % lesssim
\usepackage{graphicx} % scalebox
\usepackage{subcaption}
\usepackage{fancyhdr} % rhead; thispagestyle{fancy}
\usepackage{amsfonts}
\usepackage{tabularx} % multicolumn
\usepackage{multirow}
\usepackage{slashed}
\usepackage[
singlelinecheck=false % <-- important
]{caption}
\usepackage{feynmp} % Feynman Diagrams
\usepackage[utf8]{inputenc}

\begin{document}

\renewcommand\headrule{} % remove header line

\title{{\huge\bf{Vector-like Leptons:}}\\{\Large\bf{Muon g-2 Anomaly, Lepton Flavor Violation,\\Higgs Decays, and Lepton Non-Universality}}}
\author{Zijie Poh$^1$\footnote{poh.7@osu.edu}}
\author{Stuart Raby$^1$\footnote{raby.1@osu.edu}}
\affil{$^1$\emph{Department of Physics}\\\emph{The Ohio State University}\\\emph{191 W.~Woodruff Ave, Columbus, OH 43210, USA}}

\maketitle
\thispagestyle{fancy}
\pagenumbering{gobble} % Remove page number

%%%%%%%%%%%%%%%%%%%%%%%%%%%%%%%%%%%%%%%%%%%%%%%%%%%%%%%%%%%%%%%%%%%%%%%%%%%%%
% Abstract %%%%%%%%%%%%%%%%%%%%%%%%%%%%%%%%%%%%%%%%%%%%%%%%%%%%%%%%%%%%%%%%%%
%%%%%%%%%%%%%%%%%%%%%%%%%%%%%%%%%%%%%%%%%%%%%%%%%%%%%%%%%%%%%%%%%%%%%%%%%%%%%
\begin{abstract}\normalsize\parindent 0pt\parskip 5pt
  In this paper, we consider the Standard Model (SM) with one family of vector-like (VL) leptons, which couple to all three families of the SM leptons.  We study the constraints on this model coming from the heavy charged lepton mass bound, electroweak precision data, the muon anomalous magnetic moment, lepton flavor violation, Higgs decay constraints and a recently measured lepton non-universality observable, $R_{K^{*0}}$.  We find that the strongest constraints are coming from the muon $g-2, \ R_{\mu\mu}=\Gamma(h\to\mu\mu)/\Gamma(h\to\mu\mu)_\text{SM}, \ R_{\gamma\gamma}$ and $\text{BR}(\mu\to e\gamma)$.  Although VL leptons couple to all three families of the SM leptons, the ratio of electron-VL to muon-VL coupling is constrained to be $\langle\lambda_e/\lambda_\mu\rangle\lesssim10^{-4}$.  We also find that our model cannot fit the observed value of $R_{K^{*0}}$.
\end{abstract}

%%%%%%%%%%%%%%%%%%%%%%%%%%%%%%%%%%%%%%%%%%%%%%%%%%%%%%%%%%%%%%%%%%%%%%%%%%%%%
%%%%%%%%%%%%%%%%%%%%%%%%%%%%%%%%%%%%%%%%%%%%%%%%%%%%%%%%%%%%%%%%%%%%%%%%%%%%%
%%%%%%%%%%%%%%%%%%%%%%%%%%%%%%%%%%%%%%%%%%%%%%%%%%%%%%%%%%%%%%%%%%%%%%%%%%%%%
\pagenumbering{arabic} % Start page numbering

\newpage
\section{Introduction}
The Standard Model (SM) is a highly successful theory in predicting and fitting many experimental measurements, with few exceptions.  One of the discrepancies between the SM and experimental measurements, that has been known for a long time, is the muon anomalous magnetic moment.  The experimentally measured muon anomalous magnetic moment and the SM prediction are given by~\cite{Olive:2016xmw}
\begin{align}
  a_\mu^\text{exp} &= 11 659 209.1(5.4)(3.3)\times10^{-10} \,, \\
  a_\mu^\text{SM} &= 11 659 180.3(0.1)(4.2)(2.6)\times10^{-10} \,.
\end{align}
The discrepancy between the experimental and theoretical values is~\cite{Olive:2016xmw}
\begin{align}
  \Delta a_\mu
  = a_\mu^\text{exp}-a_\mu^\text{SM}
  = 288(63)(49)\times10^{-11} \,.
\end{align}

A simple extension of the SM that is able to explain this discrepancy is the SM with one family of VL leptons.  Derm\'i\v sek et.~al.~showed that such a model with VL leptons coupling exclusively to the muon is sufficient to explain this discrepancy~\cite{Dermisek:2013gta}.  In a more natural theory, however, the VL leptons would couple to all three families of the SM leptons, which have been studied extensively in the literature~\cite{delAguila:2008pw,Joglekar:2012vc,Kearney:2012zi,Ishiwata:2013gma}.  Due to the flavor violating nature of this model, the SM-VL couplings are known to be highly constrained.

In this paper, we try to provide a holistic point of view of the model in which the SM is extended by one family of VL leptons and the VL leptons have non-zero couplings to all three families of the SM leptons.  We are interested in the constraints on this model coming from satisfying the heavy charged lepton mass bound, electroweak precision data, the muon $g-2$, lepton flavor violation, Higgs decays and lepton non-universality observables.  We find that this model cannot simultaneously satisfy electroweak precision measurements and the $2.2-2.5\sigma$ lepton universality SM deviation in $R_{K^{*0}}=\Gamma(B^0\to K^{*0}\mu\mu)/\Gamma(B^0\to K^{*0}ee)$ measured by LHCb~\cite{Bifani:2017lhcb}.  As for the other observables, we find that the most constraining observables are the muon $g-2, \ R_{\mu\mu}=\Gamma(h\to\mu\mu)/\Gamma(h\to\mu\mu)_\text{SM}, \ R_{\gamma\gamma}$ and $\text{BR}(\mu\to e\gamma)$.

%%%%%%%%%%%%%%%%%%%%%%%%%%%%%%%%%%%%%%%%%%%%%%%%%%%%%%%%%%%%%%%%%%%%%%%%%%%%%
%%%%%%%%%%%%%%%%%%%%%%%%%%%%%%%%%%%%%%%%%%%%%%%%%%%%%%%%%%%%%%%%%%%%%%%%%%%%%
%%%%%%%%%%%%%%%%%%%%%%%%%%%%%%%%%%%%%%%%%%%%%%%%%%%%%%%%%%%%%%%%%%%%%%%%%%%%%

\section{Model}
\label{sec:ch2.model}

\begin{table}[!htbp]
  \begin{center}
    \begin{tabular}{|c|ccc|cc|}
      \hline
      & \multicolumn{3}{c|}{SM} & \multicolumn{2}{c|}{VL} \\\hline
      & $\ell_{Li}=\begin{pmatrix}\nu_{Li}\\e_{Li}\end{pmatrix}$ & $e_{Ri}$ & $H=\begin{pmatrix} \phi^+ \\v+(h + i \phi^0)/\sqrt{2}\end{pmatrix}$ & $L_{L,R}=\begin{pmatrix}L_{L,R}^0\\L_{L,R}^-\end{pmatrix}$ & $E_{L,R}$ \\\hline
      SU(2)$_L$ & 2 & 1 & 2 & 2 & 1 \\\hline
      U(1)$_Y$ & -1 & -2 & 1 & -1 & -2 \\\hline
    \end{tabular}
  \end{center}
  \caption{The quantum numbers of leptonic sector particles relevant to this paper.  $i=1,2,3$ is SM family index.  The electric charge is given by $Q=T_3+Y/2$ and the Higgs vacuum expectation value is $174\,\text{GeV}$. The fields $h, \ \phi^+, \ \phi^0$ are the physical Higgs boson and the would-be Goldstone bosons, respectively, which give the $W^\pm$ and $Z$ mass.}
  \label{tab:qm}
\end{table}

The model that we study is the SM with one generation of VL leptons.  The particles in the leptonic sector and their corresponding quantum numbers are given in Tab.~\ref{tab:qm} and the corresponding Lagrangian is given by
\begin{align}
  \begin{aligned}
    \mathcal{L} \supset&
      -\bar\ell_{Li}y_{ii}^ee_{Ri}H
      -\bar\ell_{Li}\lambda_i^EE_RH
      -\bar L_L\lambda_i^Le_{Ri}H
      -\bar L_L\lambda E_RH
      -\bar E_L\bar\lambda L_RH^\dagger
    \\&
      -M_L\bar L_LL_R
      -M_E\bar E_LE_R
      + \text{h.c.} \,,
  \end{aligned}
\end{align}
where $i=1,2,3$ is the SM family index.  Without loss of generality, we assume that the SM lepton Yukawa matrix, $y^e$, is already diagonalized.  Thus, the lepton mass matrix is
\begin{align}
  \begin{pmatrix} \bar e_{Li} & \bar L_L^- & \bar E_L \end{pmatrix}
  \begin{pmatrix}
    y_{ii}^ev & 0 & \lambda_i^Ev \\
    \lambda_i^Lv & M_L & \lambda v \\
    0 & \bar\lambda v & M_E
  \end{pmatrix}
  \begin{pmatrix} e_{Ri} \\ L_R^- \\ E_R \end{pmatrix}
  \equiv
  \bar e_{La}\mathcal{M}e_{Ra} \,,
\end{align}
where $a=1,\dots,5$.  Let $U_L$ and $U_R$ be unitary matrices that diagonalize the charged lepton mass matrix:
\begin{align}
  U_L^\dagger\mathcal{M}U_R
  =
  \begin{pmatrix}
    M_{e_i} & 0 & 0 \\
    0 & M_{e4} & 0 \\
    0 & 0 & M_{e5}
  \end{pmatrix}
  \equiv \mathcal{M}^{\text{diag}}\,.
\end{align}
and $[\hat e_{L,R}]_a=[U_{L,R}^\dagger]_{a,a'}[e_{L,R}]_{a'}$ are the mass basis.\footnote{In this model, neutrinos are assumed to only obtain a VL mass term, $M_L$.}

The $Z$-lepton couplings are given by
\begin{align}
  \mathcal L \supset \frac{g}{c_W}Z_\mu\left[\bar e_{La}\gamma^\mu(T_a^3+s_W^2)e_{La}+\bar e_{Ra}\gamma^\mu(T_a^3+s_W^2)e_{Ra}\right] \,,
\end{align}
where $s_W=\sin\theta_W, c_W=\cos\theta_W$ and $T_a^3$ is the SU(2) generator where
\begin{align}
  T_{a}^3e_{La} &= -\frac{1}{2}\text{diag}(1,1,1,1,0)e_{La} \equiv T_L^3e_{La} \\
  T_{a}^3e_{Ra} &= -\frac{1}{2}\text{diag}(0,0,0,1,0)e_{Ra} \equiv T_R^3e_{Ra} \,.
\end{align}
Since these matrices are not proportional to the identity matrix, when we rotate to the lepton mass basis, the $Z$-lepton couplings are not diagonal:
\begin{align}
  \mathcal{L} \supset Z_\mu\left[\bar{\hat e}_{La}\gamma^\mu g_{Lab}^Z\hat e_{Lb}+\bar{\hat e}_{Ra}\gamma^\mu g_{Rab}^Z\hat e_{Rb}\right] \,,
\end{align}
where $g_{L,R}^Z=(g/c_W)[U_{L,R}^\dagger(T_{L,R}^3+s_W^2)U_{L,R}]$.  Hence, this model has lepton flavor violating $Z$ boson decays.

The $W$-lepton couplings are given by
\begin{align}
  \mathcal L \supset \frac{g}{\sqrt2}W_\mu^+\left[\bar\nu_{La}\gamma^\mu e_{La} + \bar\nu_{Ra}\gamma_\mu e_{Ra}\right] + h.c. \,,
\end{align}
where
\begin{align}
  \nu_{La} = \begin{pmatrix} \nu_{Li}\\ L_L^0 \\0\end{pmatrix}
  \hspace{2cm}
  \nu_{Ra} = \begin{pmatrix} 0_i\\ L_R^0 \\0\end{pmatrix} \,.
\end{align}
Hence, in the charged lepton mass basis, we have
\begin{align}
  \mathcal L \supset W_\mu^+\left[\bar\nu_{La}\gamma^\mu g_{Lab}^W\hat e_{Lb} + \bar\nu_{Ra}\gamma_\mu g_{Rab}^W\hat e_{Rb}\right] + h.c. \,,
  \label{eq:W}
\end{align}
where $g_L^W=(g/\sqrt2)\text{diag}(1,1,1,1,0)U_L$ and $g_R^W=(g/\sqrt2)\text{diag}(0,0,0,1,0)U_R$.

The coupling between the physical Higgs boson and the leptons is
\begin{align}
  \mathcal L \supset -\frac{1}{\sqrt2}h\bar e_{La}Y_{ab}e_{Rb} + h.c. \,,
\end{align}
where
\begin{align}
  Y =
  \begin{pmatrix}
    y_{ii}^e & 0 & \lambda_i^E \\
    \lambda_i^L & 0 & \lambda \\
    0 & \bar\lambda & 0
  \end{pmatrix} \,.
\end{align}
In the mass basis, we have
\begin{align}
  \mathcal{L} \supset -\frac{1}{\sqrt2}h\bar{\hat e}_{La}\hat Y_{ab}\hat e_{Rb} + h.c. \,,
\end{align}
where
\begin{align}
  \hat Y = U_L^\dagger YU_R \,.
\end{align}
This Yukawa matrix is non-diagonal because $Yv=\mathcal M-\text{diag}(0,0,0,M_L,M_E)$.
Hence,
\begin{align}
  \hat Y = \mathcal M^\text{diag}/v-U_L^\dagger\text{diag}(0,0,0,M_L,M_E)U_R/v \,,
\end{align}
where the second term is non-diagonal.

To calculate the effect of this model on lepton non-universality, we consider the following Hamiltonian~\cite{Buras:1994dj,Bobeth:1999mk}
\begin{align}
  \mathcal H_\text{eff} = -\frac{4G_F}{\sqrt2}V_{tb}V_{ts}^*\frac{e^2}{16\pi^2}\sum_{j=9,10}C_jO_j \,,
\end{align}
where
\begin{align}
  \mathcal O_9 &= (\bar s_L\gamma^\mu b_L)(\bar{\hat e}_a\gamma_\mu\hat e_a) \,, \\
  \mathcal O_{10} &= (\bar s_L\gamma^\mu b_L)(\bar{\hat e}_a\gamma_\mu\gamma_5\hat e_a) \,.
\end{align}
The new physics (NP) contribution to these two Wilson coefficients are coming from the box diagrams in Fig.~\ref{fig:box} (see appendix for the calculation~\cite{Inami:1980fz})
\begin{align}
\begin{aligned}
  C_9^\text{NP} =& -\frac{1}{s^2_W}\frac{1}{4}\bigg[\left(|[U_L]_{4a}|^2+|[U_R]_{4a}|^2+\frac{1}{4}\frac{v^2}{M_L^2}xy(|[Y^{\nu_R}U_L]_{4a}|^2+|[Y^{\nu_L}U_R]_{4a}|^2)\right)g_1(x,y) \\
  &\hspace{3em}-\frac{v}{M_L}xy ([U_L]_{4a}[Y^{\nu_R*}U_L^*]_{4a}+[U_L^*]_{4a}[Y^{\nu_R}U_L]_{4a} \\
  &\hspace{9em}+[U_R]_{4a}[Y^{\nu_L*}U_R^*]_{4a}+[U_R^*]_{4a}[Y^{\nu_L}U_R]_{4a} )g_0(x,y)\bigg] \,,
\end{aligned}
\end{align}
\begin{align}
\begin{aligned}
  C_{10}^\text{NP} =& \frac{1}{s^2_W}\frac{1}{4}\bigg[\left(|[U_L]_{4a}|^2-|[U_R]_{4a}|^2+\frac{1}{4}\frac{v^2}{M_L^2}xy(|[Y^{\nu_R}U_L]_{4a}|^2-|[Y^{\nu_L}U_R]_{4a}|^2)\right)g_1(x,y) \\
  &\hspace{3em}-\frac{v}{M_L}xy ([U_L]_{4a}[Y^{\nu_R*}U_L^*]_{4a}+[U_L^*]_{4a}[Y^{\nu_R}U_L]_{4a} \\
  &\hspace{9em}-[U_R]_{4a}[Y^{\nu_L*}U_R^*]_{4a}-[U_R^*]_{4a}[Y^{\nu_L}U_R]_{4a} )g_0(x,y)\bigg] \,,
\end{aligned}
\end{align}
where $x=M_t^2/M_W^2, \ y=M_L^2/M_W^2$,
\begin{align}
  Y^{\nu_L} \equiv
  \begin{pmatrix}
    y_{ii}^e & 0 & \lambda_i^E \\
    \lambda_i^L & 0 & \lambda \\
    0 & 0 & 0
  \end{pmatrix}
  \text{\,,}\hspace{2em}
  Y^{\nu_R\dagger} \equiv
  \begin{pmatrix}
    0 & 0 & 0 \\
    0 & 0 & 0 \\
    0 & \bar\lambda & 0
  \end{pmatrix} \,,
\end{align}
\begin{align}
  g_1(x,y) &= \frac{1}{x-y}\left[\frac{x^2}{(x-1)^2}\log x-\frac{y^2}{(y-1)^2}\log y-\frac{1}{x-1}+\frac{1}{y-1}\right] \,, \\
  g_0(x,y) &= \frac{1}{x-y}\left[\frac{x}{(x-1)^2}\log x-\frac{y}{(y-1)^2}\log y-\frac{1}{x-1}+\frac{1}{y-1}\right] \,.
\end{align}

\begin{figure}
\captionsetup[subfigure]{justification=centering}
\begin{center}
\begin{subfigure}[b]{0.4\textwidth}
  \begin{center}
  \begin{fmffile}{box_diag_a}
    \begin{fmfgraph*}(100, 70)
      \fmfstraight
      \fmfleft{l1,l2,l3,l4}
      \fmfright{r1,r2,r3,r4}
      \fmf{phantom}{l4,w1,w2,w3,r4}
      \fmf{phantom}{l3,v7,v8,v9,r3}
      \fmf{phantom}{l2,v4,v5,v6,r2}
      \fmf{phantom}{l1,v1,v2,v3,r1}
      \fmffreeze
      \fmf{fermion}{l1,v1}
      \fmf{fermion}{v3,r1}
      \fmf{fermion,label=$u,,c,,t$}{v1,v3}
      \fmf{fermion}{v7,w2}
      \fmf{fermion}{r4,v9}
      \fmf{fermion,label=$\nu_4$,label.side=left}{v9,v7}
      \fmflabel{$b$}{l1}
      \fmflabel{$s$}{r1}
      \fmflabel{$\hat e_a$}{r4}
      \fmflabel{$\hat e_a$}{w2}
      \fmf{boson,label=$W^-$,label.side=left}{v1,v7}
      \fmf{boson,label=$W^+$}{v3,v9}
    \end{fmfgraph*}
  \end{fmffile}
  \vspace{1em}
  \caption{}
  \end{center}
\end{subfigure}
\hspace{4em}
\begin{subfigure}[b]{0.4\textwidth}
  \begin{center}
  \begin{fmffile}{box_diag_b}
    \begin{fmfgraph*}(100, 70)
      \fmfstraight
      \fmfleft{l1,l2,l3,l4}
      \fmfright{r1,r2,r3,r4}
      \fmf{phantom}{l4,w1,w2,w3,r4}
      \fmf{phantom}{l3,v7,v8,v9,r3}
      \fmf{phantom}{l2,v4,v5,v6,r2}
      \fmf{phantom}{l1,v1,v2,v3,r1}
      \fmffreeze
      \fmf{fermion}{l1,v1}
      \fmf{fermion}{v3,r1}
      \fmf{fermion,label=$u,,c,,t$}{v1,v3}
      \fmf{fermion}{v7,w2}
      \fmf{fermion}{r4,v9}
      \fmf{fermion,label=$\nu_4$,label.side=left}{v9,v7}
      \fmflabel{$b$}{l1}
      \fmflabel{$s$}{r1}
      \fmflabel{$\hat e_a$}{r4}
      \fmflabel{$\hat e_a$}{w2}
      \fmf{dashes,label=$\phi^-$,label.side=left}{v1,v7}
      \fmf{boson,label=$W^+$}{v3,v9}
    \end{fmfgraph*}
  \end{fmffile}
  \vspace{1em}
  \caption{}
  \end{center}
\end{subfigure}
\\\vspace{2em}
\begin{subfigure}[b]{0.4\textwidth}
  \begin{center}
  \begin{fmffile}{box_diag_c}
    \begin{fmfgraph*}(100, 70)
      \fmfstraight
      \fmfleft{l1,l2,l3,l4}
      \fmfright{r1,r2,r3,r4}
      \fmf{phantom}{l4,w1,w2,w3,r4}
      \fmf{phantom}{l3,v7,v8,v9,r3}
      \fmf{phantom}{l2,v4,v5,v6,r2}
      \fmf{phantom}{l1,v1,v2,v3,r1}
      \fmffreeze
      \fmf{fermion}{l1,v1}
      \fmf{fermion}{v3,r1}
      \fmf{fermion,label=$u,,c,,t$}{v1,v3}
      \fmf{fermion}{v7,w2}
      \fmf{fermion}{r4,v9}
      \fmf{fermion,label=$\nu_4$,label.side=left}{v9,v7}
      \fmflabel{$b$}{l1}
      \fmflabel{$s$}{r1}
      \fmflabel{$\hat e_a$}{r4}
      \fmflabel{$\hat e_a$}{w2}
      \fmf{boson,label=$W^-$,label.side=left}{v1,v7}
      \fmf{dashes,label=$\phi^+$}{v3,v9}
    \end{fmfgraph*}
  \end{fmffile}
  \vspace{1em}
  \caption{}
  \end{center}
\end{subfigure}
\hspace{4em}
\begin{subfigure}[b]{0.4\textwidth}
  \begin{center}
  \begin{fmffile}{box_diag_d}
    \begin{fmfgraph*}(100, 70)
      \fmfstraight
      \fmfleft{l1,l2,l3,l4}
      \fmfright{r1,r2,r3,r4}
      \fmf{phantom}{l4,w1,w2,w3,r4}
      \fmf{phantom}{l3,v7,v8,v9,r3}
      \fmf{phantom}{l2,v4,v5,v6,r2}
      \fmf{phantom}{l1,v1,v2,v3,r1}
      \fmffreeze
      \fmf{fermion}{l1,v1}
      \fmf{fermion}{v3,r1}
      \fmf{fermion,label=$u,,c,,t$}{v1,v3}
      \fmf{fermion}{v7,w2}
      \fmf{fermion}{r4,v9}
      \fmf{fermion,label=$\nu_4$,label.side=left}{v9,v7}
      \fmflabel{$b$}{l1}
      \fmflabel{$s$}{r1}
      \fmflabel{$\hat e_a$}{r4}
      \fmflabel{$\hat e_a$}{w2}
      \fmf{dashes,label=$\phi^-$,label.side=left}{v1,v7}
      \fmf{dashes,label=$\phi^+$}{v3,v9}
    \end{fmfgraph*}
  \end{fmffile}
  \vspace{1em}
  \caption{}
  \end{center}
\end{subfigure}
  \caption{Box diagrams contributing to $b\to s\hat e_a\hat e_a$.}
  \label{fig:box}
\end{center}
\end{figure}
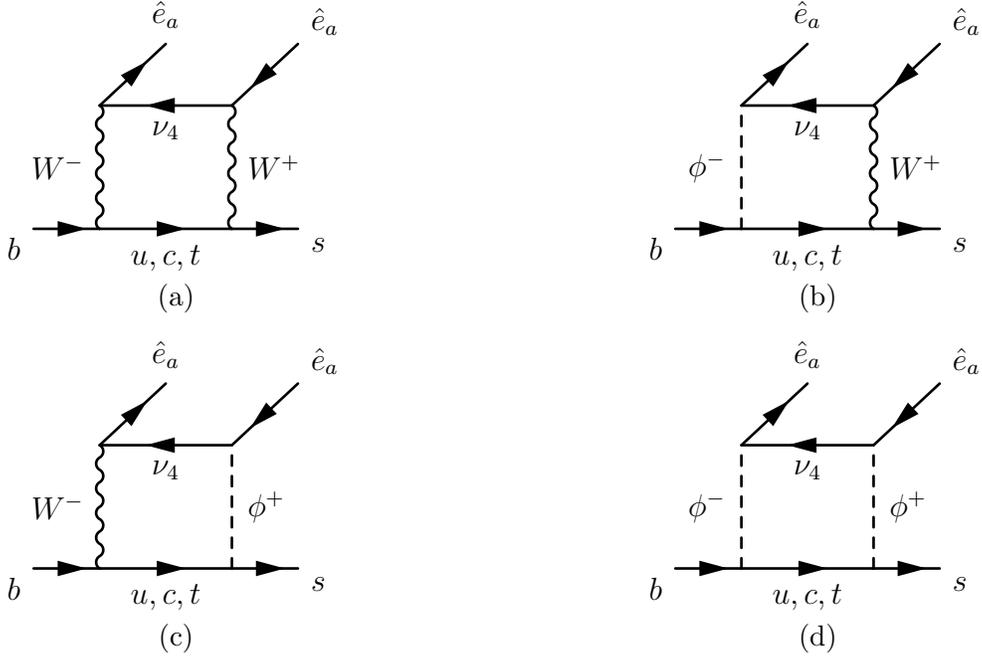

%%%%%%%%%%%%%%%%%%%%%%%%%%%%%%%%%%%%%%%%%%%%%%%%%%%%%%%%%%%%%%%%%%%%%%%%%%%%%
%%%%%%%%%%%%%%%%%%%%%%%%%%%%%%%%%%%%%%%%%%%%%%%%%%%%%%%%%%%%%%%%%%%%%%%%%%%%%
%%%%%%%%%%%%%%%%%%%%%%%%%%%%%%%%%%%%%%%%%%%%%%%%%%%%%%%%%%%%%%%%%%%%%%%%%%%%%

\section{Procedure}

The analysis of this paper is similar to that in~\cite{Dermisek:2013gta}.  A new feature of this paper is that we do not assume that VL leptons couple exclusively to muons.  Instead, we allow non-zero SM-VL leptons coupling and are interested in the constraints of the 10 model parameters: $M_{L,E}, \ \lambda, \ \bar\lambda$ and $\lambda_{e,\mu,\tau}^{L,E}$.  $y_{e,\mu,\tau}$ are not free parameters because we choose $y_{e,\mu,\tau}$ such that $m_{e,\mu,\tau}$ are the central values in Ref.~\cite{Olive:2016xmw}.  We considered $M_{L,E}\in(100,1000)\,\text{GeV}$ and $\lambda,\bar\lambda\in(-1,1)$.  As for the SM-VL couplings, we considered
\begin{align}
  \frac{\lambda_{e,\mu,\tau}^{L,E}v}{M_{L,E}}\in(-0.09,0.09) \,.
\end{align}
The ranges of the SM-VL couplings are chosen to satisfy the electroweak constraints.

The constraints that we consider in this paper are from the heavy charged lepton mass bound, electroweak precision data, the muon $g-2$, lepton flavor violation, Higgs decay and a lepton non-universality observable, $R_{K^{*0}}$. See Tab.~\ref{tab:obs} for the complete list of observables.  All of the experimental values, other than $R_{K^{*0}}$, are taken from Ref.~\cite{Olive:2016xmw}.  $R_{K^{*0}}$ is recently measured by LHCb~\cite{Bifani:2017lhcb}.  All theoretical calculations are performed at leading order, that is all observables other than $\Delta a_\mu, \ \text{BR}(\ell\to\ell'\gamma)$ and $R_{\gamma\gamma}$ are calculated at tree level.  The theoretical calculation of the VL contribution to the muon $g-2$ is taken from Ref.~\cite{Dermisek:2013gta}.  The calculation for $\text{BR}(\ell\to\ell'\gamma)$ and $R_{\gamma\gamma}$ are performed at one-loop level~\cite{Lavoura:2003xp,Djouadi:2005gi}.  Since all calculations are performed at leading order, we have included 1\% theoretical error when ensuring that the calculated observables satisfy the current experimental bounds.  As for lepton non-universality analysis, we have used \texttt{flavio}, a very versatile program that calculates $b$-physics observables written by Straub et.~al.~\cite{david_straub_2017_569011}.  To calculate the NP effects of the observables implemented in \texttt{flavio}, one only has to specify the NP contribution to the Wilson coefficients.

 In our analysis we obtain a scatter plot by sampling from the parameter space and checking to see if the sampled points satisfy the constraints mentioned above.  To ensure that we cover all regions in this vast parameter space, we divide VL masses into four different regions: $M_{L,E}\in[100,150),[150,250),[250,500),[500,1000)\,\text{GeV}$, and the VL-VL couplings into two different regions\footnote{These couplings can be positive or negative.  The quoted ranges are the magnitude.  Similarly for SM-VL couplings.}: $|\lambda|, |\bar\lambda|\in[0,0.75),[0.75,1)$.  As for the muon-VL coupling, we considered $|\lambda_\mu^{L,E}v/M_{L,E}|\in[0,0.06),[0.06,0.09)$.  For each of these regions, we sampled 10,000 points satisfying the heavy charged lepton mass bound and the electroweak precision observables.  The total number of simulated points is 2.56 millions points.  The parameters $M_{L,E}, \ \lambda, \ \bar{\lambda}, \ \lambda_\mu^{L,E}$ are sampled from a uniform distribution while $|\lambda_{e,\tau}^{L,E}v/M_{L,E}|\in[10^{-10},0.09)$ are sampled from a log-uniform distribution.  The electron-VL and tau-VL couplings are sampled from a log-uniform distribution because we expect these couplings to be highly constrained by flavor violation observables and we are interested in determining the degree of fine-tuning in these two parameters in order to be consistent with the flavor violation constraints.

\begin{table}[!htbp]
  \begin{center}
    \begin{tabular}{|c|c|c|}
      \hline
      Muon $g-2$ & $\mu$ & $\Delta a_\mu$ \\\hline
      Heavy Charged Leptons & $e_4$ & $M_{e4}$ \\\hline
      \multirow{5}{*}{\shortstack{Electroweak\\Precision}} & \multirow{2}{*}{$Z$} & $A_{e,\mu,\tau},A_{FB}^{(0e),(0\mu),(0\tau)}$ \\
      & & $\text{BR}(Z\to ee),\text{BR}(Z\to\mu\mu),\text{BR}(Z\to\tau\tau)$ \\\cline{2-3}
      & $W$ & $\text{BR}(W\to e\nu_e),\text{BR}(W\to\mu\nu_\mu),\text{BR}(W\to\tau\nu_\tau)$ \\\cline{2-3}
      & $\mu$ & $\text{BR}(\mu\to e\bar\nu_e\nu_\mu)$ \\\cline{2-3}
      & $\tau$ & $\text{BR}(\tau\to e\bar\nu_e\nu_\tau),\text{BR}(\tau\to\mu\bar\nu_e\nu_\tau)$ \\\hline
      \multirow{4}{*}{\shortstack{Lepton\\Flavor\\Violation}} & $Z$ & $\text{BR}(Z\to e\mu),\text{BR}(Z\to e\tau),\text{BR}(Z\to\mu\tau)$ \\\cline{2-3}
      & $\mu$ & $\text{BR}(\mu\to e\gamma),\text{BR}(\mu\to 3e)$ \\\cline{2-3}
      & \multirow{2}{*}{$\tau$} & $\text{BR}(\tau\to e\gamma),\text{BR}(\tau\to\mu\gamma)$ \\
      & & $\text{BR}(\tau\to 3e),\text{BR}(\tau\to3\mu)$ \\\hline
      Higgs & $h$ & $R_{\mu\mu}, R_{\tau\tau}, R_{\gamma\gamma}, \text{BR}(h\to\mu\tau)$ \\\hline
      Lepton Non-Universality & $B^0$ & $R_{K^{*0}}$ \\\hline
    \end{tabular}
  \end{center}
  \caption{The complete list of observables considered in this paper.  $\Delta a_\mu$ is the difference between the measured muon $g-2$ and the SM prediction.  $A_{e,\mu,\tau}$ is the electron, muon and tau left-right asymmetry in Z decay.  $A_{FB}^{(0e),(0\mu),(0\tau)}$ is the electron, muon and tau forward-backward asymmetry in Z decay.  $R_{\mu\mu}=\Gamma(h\to\mu\mu)/\Gamma(h\to\mu\mu)_\text{SM}$ and similarly for $R_{\tau\tau}$ and $R_{\gamma\gamma}$.  $R_{K^{*0}}=\Gamma(B^0\to K^{*0}\mu\mu)/\Gamma(B^0\to K^{*0}ee)$.  All experimental values, other than $R_{K^{*0}}$, are taken from Ref.~\cite{Olive:2016xmw}.  $R_{K^{*0}}$ is taken from the most recent LHCb measurement~\cite{Bifani:2017lhcb}.}
  \label{tab:obs}
\end{table}

%%%%%%%%%%%%%%%%%%%%%%%%%%%%%%%%%%%%%%%%%%%%%%%%%%%%%%%%%%%%%%%%%%%%%%%%%%%%%
%%%%%%%%%%%%%%%%%%%%%%%%%%%%%%%%%%%%%%%%%%%%%%%%%%%%%%%%%%%%%%%%%%%%%%%%%%%%%
%%%%%%%%%%%%%%%%%%%%%%%%%%%%%%%%%%%%%%%%%%%%%%%%%%%%%%%%%%%%%%%%%%%%%%%%%%%%%

\section{Results}

\begin{figure}
  \begin{center}
    \includegraphics[width=0.9\textwidth]{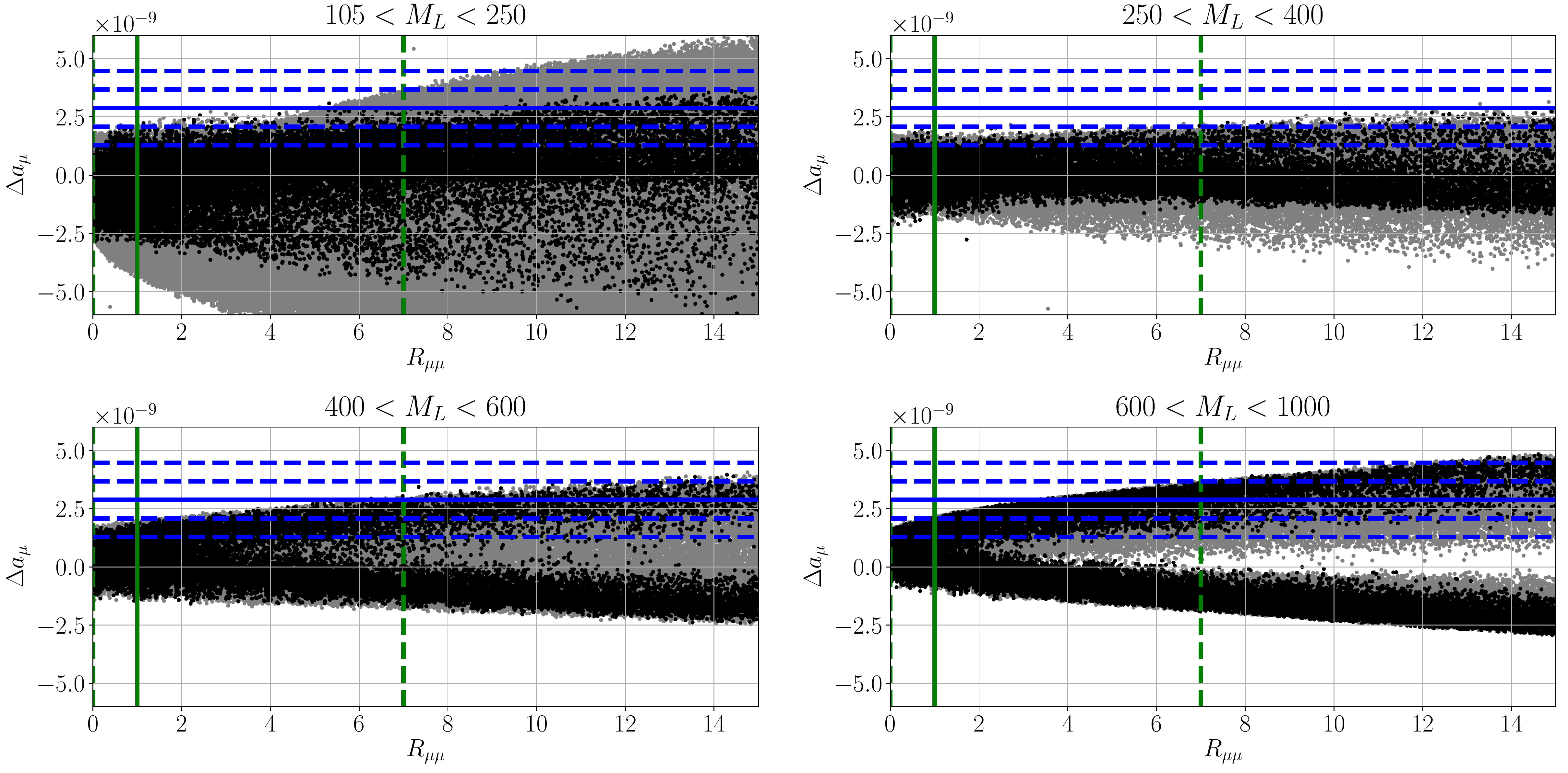}
    \caption{Plots of muon $g-2$ discrepancy, $\Delta a_\mu$, versus $R_{\mu\mu}=\Gamma(h\to\mu\mu)/\Gamma(h\to\mu\mu)_\text{SM}$.  The four plots are for different ranges of $M_L$.  The gray points failed one or more LFV and Higgs decay observables, other than $R_{\mu\mu}$, while the black points satisfy all of these observables.  The blue dashed lines are the 1 and $2\sigma$ bounds of $\Delta a_\mu$ and the green dashed line is the upper bound of $R_{\mu\mu}$.  The blue solid line is the central value of $\Delta a_\mu$ while the green solid line is $R_{\mu\mu}=1$.}
    \label{fig:mg2_vs_Rmumu}
  \end{center}
\end{figure}

In Fig.~\ref{fig:mg2_vs_Rmumu}, we plotted $\Delta a_\mu$ versus $R_{\mu\mu}$.  The four plots in this figure are for different ranges of $M_L$.  $M_L$ is a meaningful discriminator because the VL contribution to thr muon $g-2$ from the $W$-boson loop is due to the SU(2) doublet VL neutrinos, $L_{L,R}^0$, which has mass $M_L$~\cite{Dermisek:2013gta}.  The gray points do not satisfy one or more LFV and Higgs decay observables, other than $R_{\mu\mu}$.  See Tab.~\ref{tab:obs} for the complete list of observables.  On the other hand, the black points satisfy all the LFV and Higgs decay observables other than $R_{\mu\mu}$.  The blue dashed lines are the 1 and $2\sigma$ bounds of $\Delta a_\mu$ and the green dashed line is the upper bound of $R_{\mu\mu}$.  The blue solid line is the central value of $\Delta a_\mu$ while the green solid line is $R_{\mu\mu}=1$.  Notice that there are no measurement on $R_{\mu\mu}$ yet.  There is only an upper bound of $R_{\mu\mu}$.  From this figure, we see that this model can be ruled out in the future if future measurements of the muon $g-2$ and $R_{\mu\mu}$ have much smaller uncertainties, and $R_{\mu\mu}$ is measured to be SM-like while the muon $g-2$ is measured to have a similar central value.

\begin{figure}
  \begin{center}
    \includegraphics[width=0.9\textwidth]{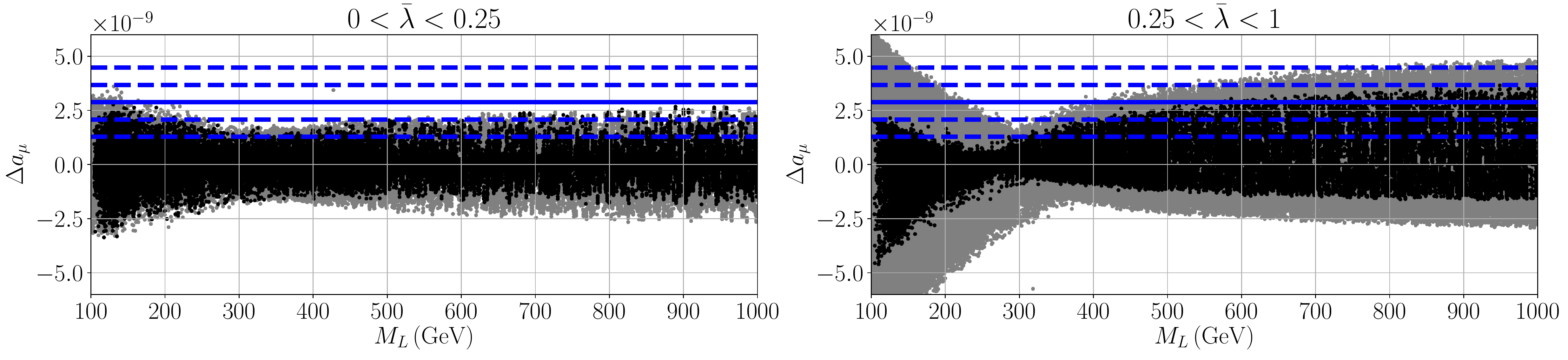}
    \caption{Plots of $\Delta a_\mu$ versus $M_L$.  The two plots are for different regions of $\bar\lambda$.  The gray points failed one or more LFV and Higgs decay observables listed in Tab.~\ref{tab:obs} while the black points satisfy all of these observables.  The dashed lines are the 1 and $2\sigma$ bounds of $\Delta a_\mu$ while the solid horizontal line is the central value of $\Delta a_\mu$.}
    \label{fig:mg2_vs_ML}
  \end{center}
\end{figure}

Fig.~\ref{fig:mg2_vs_Rmumu} also shows that the there are no points with $250\,\text{GeV}<M_L<400\,\text{GeV}$ that fits the muon $g-2$ within $1\sigma$ uncertainty\footnote{The bounds on parameter space that we obtain from this analysis are not strict because of our analysis method.  We perform the analysis by random sampling in this vast parameter space.  Our sampling method attempts to cover the parameter space as uniformly as possible.  However, we want to point out that there might still be regions of parameters space that might be missed by our sampling method.}.  This observation is further illustrated in Fig.~\ref{fig:mg2_vs_ML}, where we have plotted $\Delta a_\mu$ versus $M_L$.  The two plots in this figure are for different ranges of $\bar\lambda$.  The gray points do not satisfy one or more LFV and Higgs decay observables listed in Tab.~\ref{tab:obs} while the black points satisfy all of these observables.  Fig.~\ref{fig:mg2_vs_ML} shows that for $\bar\lambda<0.25$, this model requires $M_L<250\,\text{GeV}$ or $M_L>600\,\text{GeV}$ to fit the muon $g-2$ within $1\sigma$.  On the other hand, for $\bar\lambda>0.25$, this model requires $M_L>400\,\text{GeV}$.  Also illustrated in this plot is that the allowed region of parameter space for $M_L\lesssim250\,\text{GeV}$ can potentially be eliminated by the upcoming Fermilab E989 experiment if the central value stays the same while the uncertainty decreases by a couple factors~\cite{Chapelain:2017syu}.

\begin{figure}
  \begin{center}
    \includegraphics[width=0.9\textwidth]{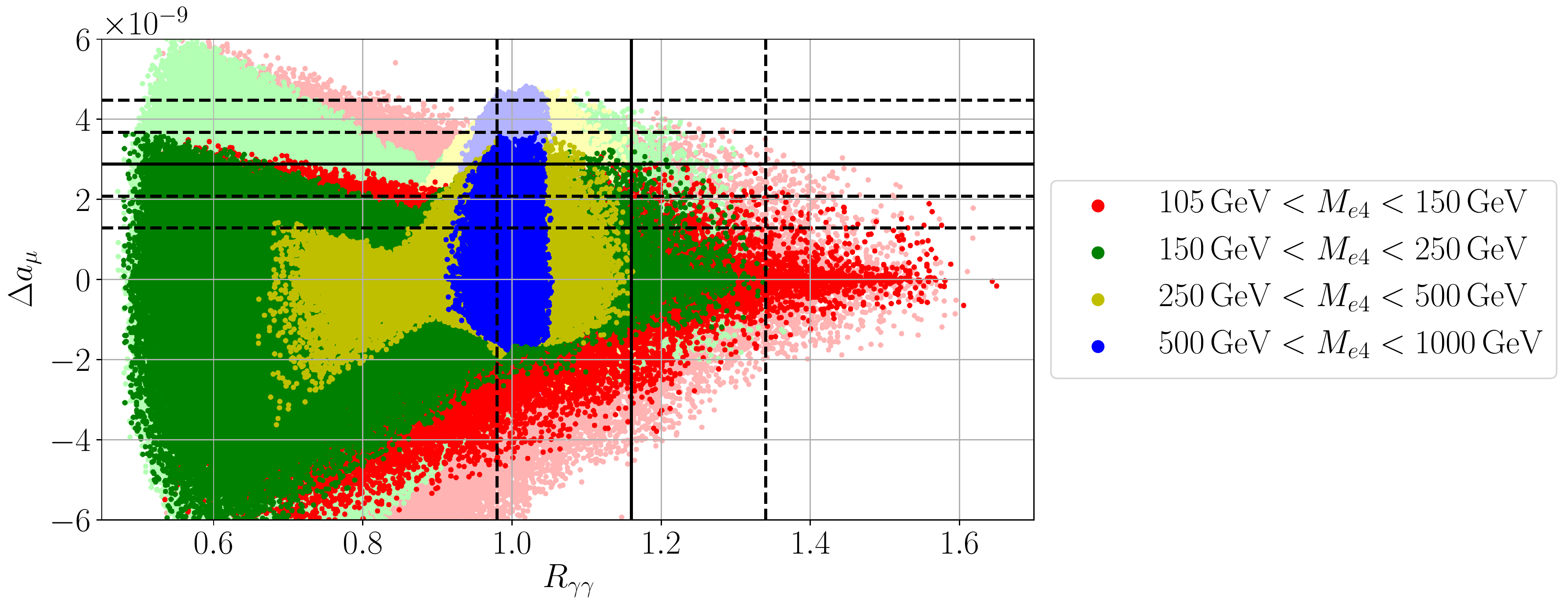}
    \caption{Plot of $\Delta a_\mu$ versus $R_{\gamma\gamma}$.  The lightly shaded points failed one or more LFV and Higgs decay observables, other than $R_{\gamma\gamma}$ while the solid colored points satisfy all of these observables.  The black dashed horizontal lines are the 1 and $2\sigma$ bounds of $\Delta a_\mu$ while the black dashed vertical lines are the $1\sigma$ bound of $R_{\gamma\gamma}$.  The black solid horizontal line is the central value of $\Delta a_\mu$ while the black solid vertical line is that of $R_{\gamma\gamma}$.}
    \label{fig:mg2_vs_Rgg}
  \end{center}
\end{figure}

In Fig.~\ref{fig:mg2_vs_Rgg}, we plotted $\Delta a_\mu$ versus $R_{\gamma\gamma}$.  The light colored points do not satisfy one or more LFV and Higgs decay observables, other than $R_{\gamma\gamma}$, while the solid colored points satisfy all the LFV and Higgs decay observables other than $R_{\gamma\gamma}$.  The ranges of the colored points are identical to that in Fig.~\ref{fig:mg2_vs_Rmumu}.  However, the points in this plot are separated into different colors based on $M_{e4}$ instead of $M_L$.  $M_{e4}$ is more meaningful in this plot because the VL leptons running in the loop of $h\to\gamma\gamma$ are the VL mass eigenstates.  As expected, for heavier VL mass eigenstates $R_{\gamma\gamma}$ is clustered around $1$.  From this plot, we learn that $M_{e4}>500\,\text{GeV}$ is a more robust region than regions with smaller $M_{e4}$ because a larger percentage of simulated points are within the experimental bound.  A very interesting scenario will arise if the central value of $R_{\gamma\gamma}$ stays and uncertainties in the measurement decrease as more data are collected.  In this scenario, we will have the potential to place an upper bound on the mass of the lightest VL mass eigenstate because there are no points with $M_{e4}>500\,\text{GeV}$ and $R_{\gamma\gamma}\gtrsim1.1$.

\begin{figure}
  \begin{center}
    \includegraphics[width=0.9\textwidth]{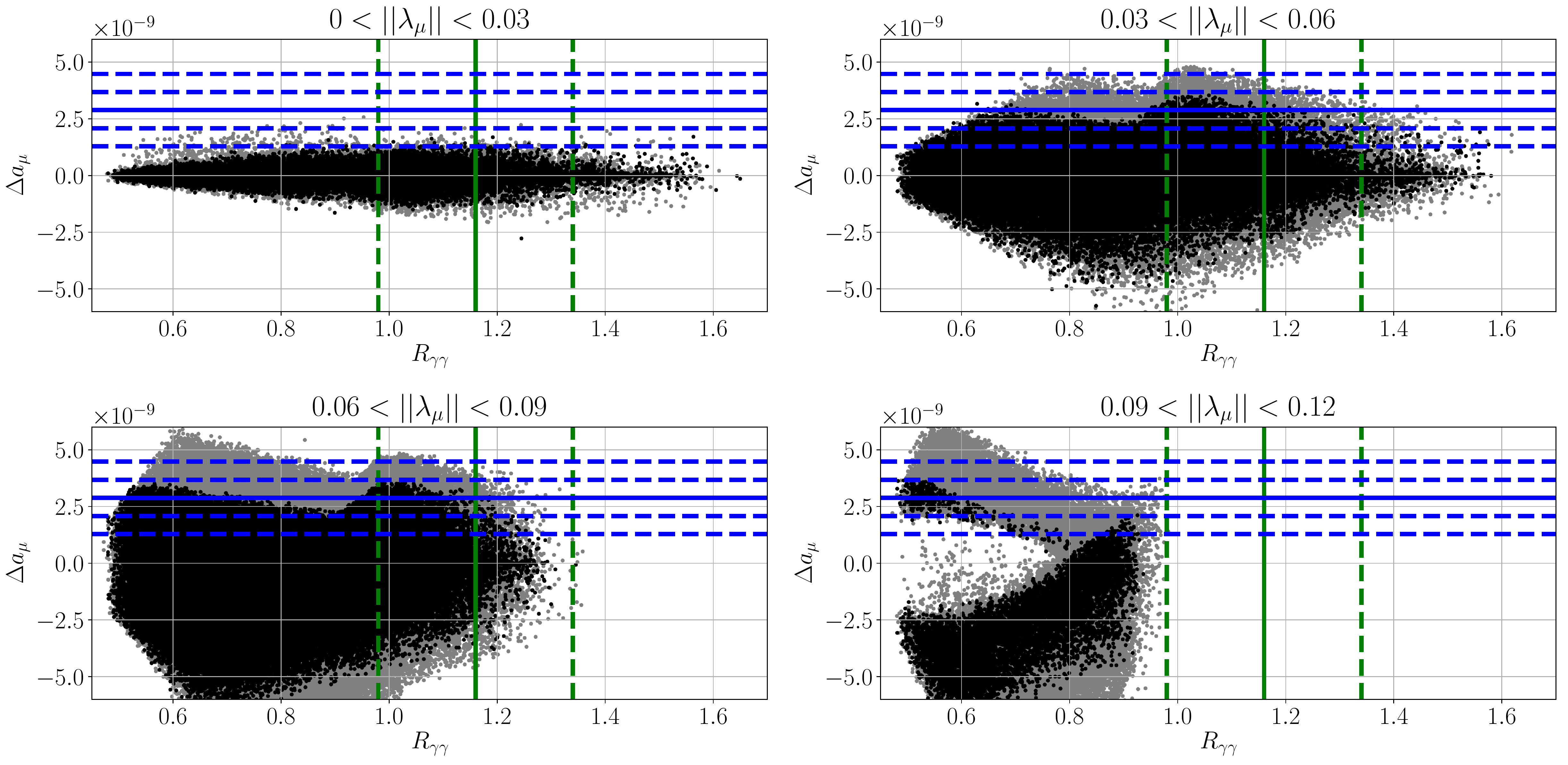}
    \caption{This figure is the same as Fig.~\ref{fig:mg2_vs_Rgg} other than the sampled points are separated into four different plots based on different values of $||\lambda_\mu||$, defined in Eq.~\ref{eq:lLEmu}.  The blue dashed lines are the 1 and $2\sigma$ bounds of $\Delta a_\mu$ and the green dashed lines are the $1\sigma$ bound of $R_{\gamma\gamma}$.  The blue solid line is the central value of $\Delta a_\mu$ while the green solid line is that of $R_{\gamma\gamma}$.  This figure shows that this model requires $||\lambda_\mu||>0.03$ to fit $\Delta_\mu$ and $||\lambda_\mu||<0.09$ to fit $R_{\gamma\gamma}$.}
    \label{fig:mg2_vs_Rgg_lLEmu}
  \end{center}
\end{figure}

Fig.~\ref{fig:mg2_vs_Rgg_lLEmu} is identical to Fig.~\ref{fig:mg2_vs_Rgg} other than the sampled points are separated into four different plots based on different values of
\begin{align}
  ||\lambda_\mu|| \equiv \sqrt{\left(\frac{\lambda_\mu^Lv}{M_L}\right)^2+\left(\frac{\lambda_\mu^Ev}{M_E}\right)^2}
  \label{eq:lLEmu}
\end{align}
instead of $M_{e4}$.  $||\lambda_\mu||$ is a meaningful variable because muon-VL coupling plays a significant role in fitting $\Delta a_\mu$ and this variable captures the norm of the muon-VL coupling normalized by the VL masses.  From this figure, we see that this model requires $||\lambda_\mu||>0.03$ to fit $\Delta a_\mu$ within $1\sigma$ and $||\lambda_\mu||<0.09$ to fit $R_{\gamma\gamma}$.

\begin{figure}
  \begin{center}
    \includegraphics[width=0.9\textwidth]{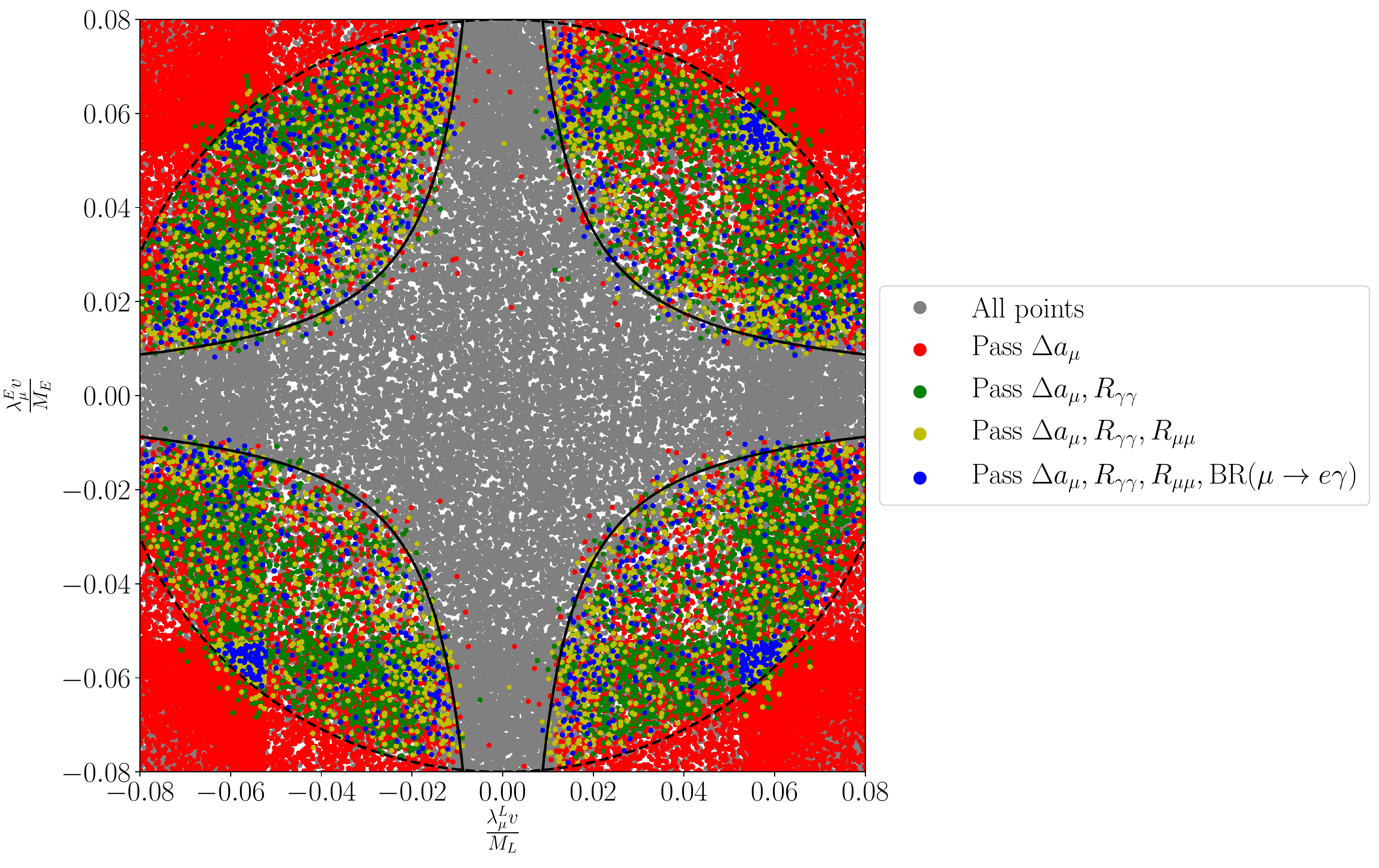}
    \caption{Plot of the muon-VL couplings, $\lambda_\mu^Ev/M_E$ versus $\lambda_\mu^Lv/M_L$.  The solid and dashed black lines are the approximate empirical bounds on the muon-VL couplings.  These bounds are not exact, but are obtained empirically (see text for more discussions).}
    \label{fig:lEmu_vs_lLmu}
  \end{center}
\end{figure}

Fig.~\ref{fig:lEmu_vs_lLmu} shows a plot of $\lambda_\mu^L$ versus $\lambda_\mu^E$.  The gray points satisfy the heavy charged lepton mass bound and electroweak precision observables; the red points satisfy the preceding constraint and $\Delta a_\mu$; the green points satisfy the preceding constraints and $R_{\gamma\gamma}$; the yellow points satisfy the preceding constraints and $R_{\mu\mu}$; the blue points satisfy all constraints listed in Tab.~\ref{tab:obs}, other than $R_{K^{*0}}$.  All the cuts are made based on the $1\sigma$ bound of the corresponding observables.  This figure shows that to satisfy $\Delta a_\mu$, the muon-VL coupling needs to satisfy approximately the following condition:
\begin{align}
  \left|\frac{\lambda_\mu^Ev}{M_E}\frac{\lambda_\mu^Lv}{M_L}\right| \gtrsim 7\times10^{-4} \,,
  \label{eq:lLEmu_constraint1}
\end{align}
which is shown by the solid lines in the figure.  It is important to notice that this bound is not an exact bound but an empirically obtained bound by requiring most simulated points to satisfy $\Delta a_\mu$ within $1\sigma$.  On the other hand, to satisfy both $\Delta a_\mu$ and $R_{\gamma\gamma}$, the muon-VL coupling needs to satisfy approximately the following condition:
\begin{align}
  \left(\frac{\lambda_\mu^Ev}{M_E}\right)^2 + \frac{1}{1.08}\left(\frac{\lambda_\mu^Lv}{M_L}\right)^2 \lesssim 0.08^2 \,,
\end{align}
which is showed by the dashed black lines.  Similar to Eq.~\ref{eq:lLEmu_constraint1}, this is not an exact bound.

\begin{figure}
  \begin{center}
    \includegraphics[width=0.9\textwidth]{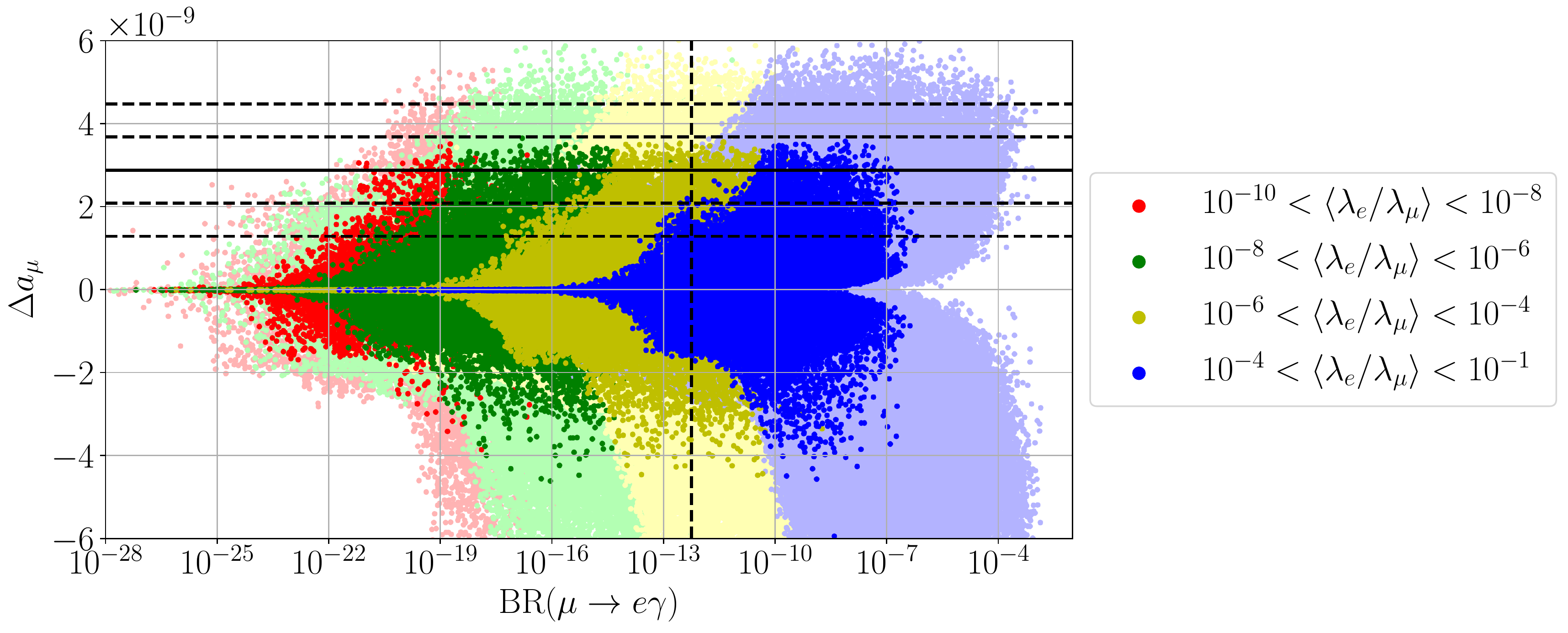}
    \caption{Figure of $\Delta a_\mu$ versus $\text{BR}(\mu\to e\gamma)$, which is the strongest LFV constraint.  The light colored points do not satisfy one or more Higgs decay and LFV observables, other than $\text{BR}(\mu\to e\gamma)$, while the solid colored points satisfy all the above mentioned constraints.  The black dashed horizontal lines are the 1 and $2\sigma$ bounds of $\Delta a_\mu$ while the black dashed vertical line is the upper bound of $\text{BR}(\mu\to e\gamma)$.  The black solid horizontal line is the central value of $\Delta a_\mu$.  This figure shows that simultaneously satisfying $\Delta a_\mu$ to within $1\sigma$  and $\text{BR}(\mu\to e\gamma)$ requires $\langle\lambda_e/\lambda_\mu\rangle<10^{-4}$.}
    \label{fig:mg2_vs_meg}
  \end{center}
\end{figure}

Fig.~\ref{fig:mg2_vs_meg} shows $\Delta a_\mu$ versus $\text{BR}(\mu\to e\gamma)$, which gives the strongest LFV constraint.  The light colored points do not satisfy one or more Higgs decay and LFV observables, other than $\text{BR}(\mu\to e\gamma)$, while the solid colored points satisfy all Higgs decay and LFV observables other than $\text{BR}(\mu\to e\gamma)$.  The sampled points in this figure are separated into four different colors based on different values of the ratio of electron-VL to muon-VL coupling:
\begin{align}
  \left\langle\frac{\lambda_e}{\lambda_\mu}\right\rangle
  \equiv \frac{1}{2}\left(\frac{\lambda_e^L}{\lambda_\mu^L}+\frac{\lambda_e^E}{\lambda_\mu^E}\right) \,.
\end{align}
The reason for separating the sampled points with this ratio is to illustrate the fine-tuning of the electron-VL coupling to the muon-VL couplings in order to satisfy LFV constraints.  The black dashed horizontal lines are the 1 and $2\sigma$ bounds of $\Delta a_\mu$ while the black dashed vertical line is the upper bound of $\text{BR}(\mu\to e\gamma)$.  The black solid horizontal line is the central value of $\Delta a_\mu$.  This figure shows that simultaneously satisfying $\Delta a_\mu$ to within 1 $\sigma$ and $\text{BR}(\mu\to e\gamma)$ requires $\langle\lambda_e/\lambda_\mu\rangle\lesssim10^{-4}$.  Again, this bound is not an exact bound\footnote{Out of all our the 2.56 million simulated points, there are 4 points that this bound does not apply to.  However, the largest value of this ratio is $\langle\lambda_e/\lambda_\mu\rangle=2\times10^{-4}$.}.

The most stringent constraints for the tau-VL coupling is coming from electroweak observables.  The range that we sample the tau-VL coupling, $\lambda_\tau^{L,E}v/M_{L,E}\in(-0.09,0.09)$, is based on electroweak constraints.  The next strongest constraints for the tau-VL coupling is $\text{BR}(\tau\to\mu\gamma)$.  However, this constraint does not rule out any value of $\lambda_\tau^{L,E}v/M_{L,E}$ within the sampling range.  Finally, the $\text{BR}(h\to\mu\tau)$ does not constrain the parameter space at all.

\begin{figure}
  \begin{center}
    \includegraphics[width=0.9\textwidth]{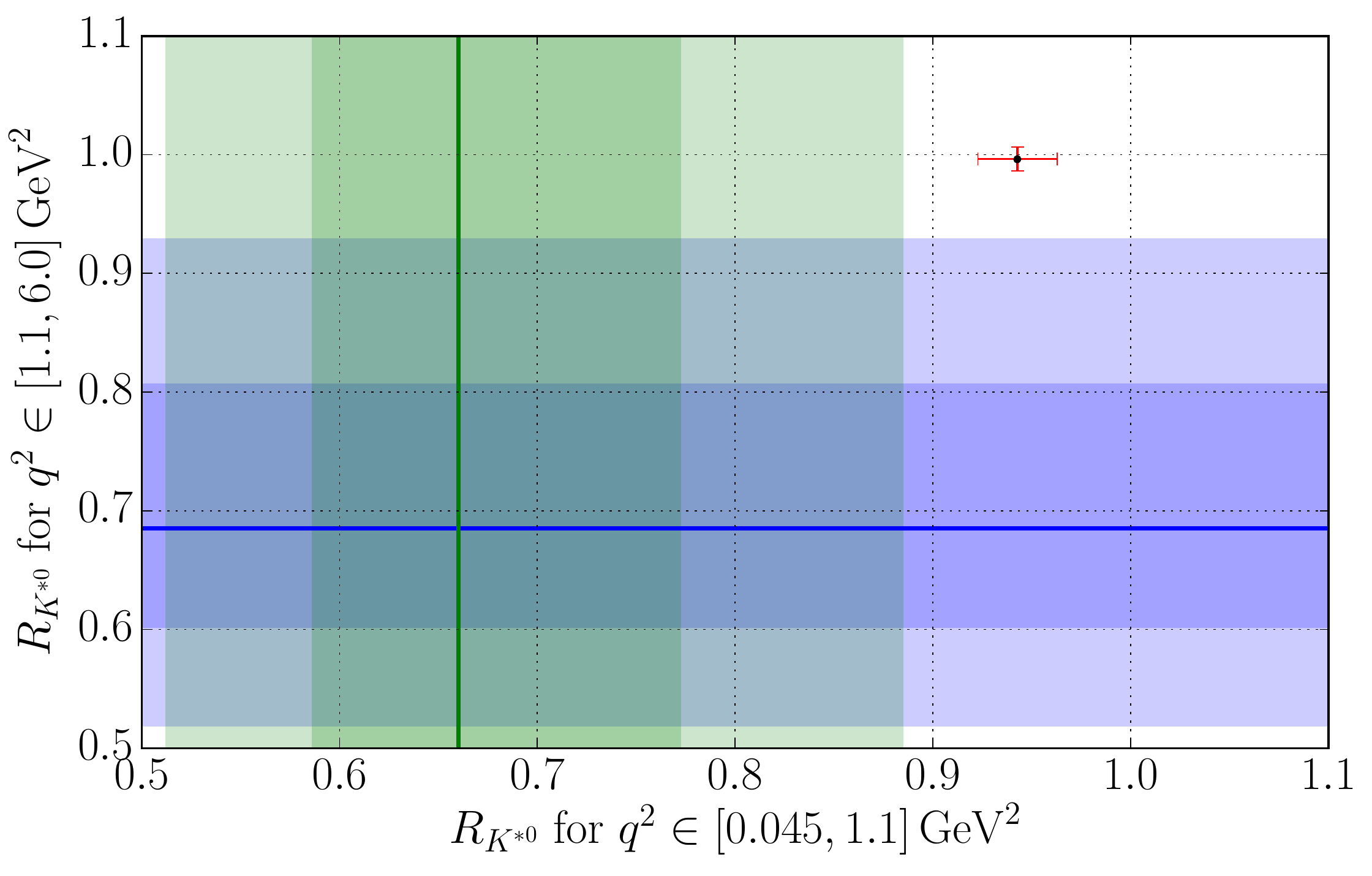}
    \caption{Plot of $R_{K^{*0}}$ for $q^2\in[0.045,1.1]\,\text{GeV}^2$ versus that for $q^2\in[1.1,6.0]\,\text{GeV}^2$.  The green bands are the 1 and $2\sigma$ uncertainty of the measured $R_{K^{*0}}$ for $q^2\in[0.045,1.1]\,\text{GeV}^2$ while the blue bands are that for $q^2\in[1.1,6.0]\,\text{GeV}^2$.  The red error bar is the SM uncertainty while the black dots are the points for this model that passes all constraints listed in Tab.~\ref{tab:obs}, other than $R_{K^{*0}}$.  This figure shows that our model cannot fit $R_{K^{*0}}$.}
    \label{fig:RKs}
  \end{center}
\end{figure}

Fig.~\ref{fig:RKs} shows the plots of $R_{K^{*0}}$ for $q^2\in[1.1,6.0]\,\text{GeV}^2$ versus that for $q^2\in[0.045,1.1]\,\text{GeV}^2$.  The green bands are the 1 and $2\sigma$ uncertainty of the measured $R_{K^{*0}}$ for $q^2\in[0.045,1.1]\,\text{GeV}^2$ while the blue bands are that for $q^2\in[1.1,6.0]\,\text{GeV}^2$.  The red error bar is the SM uncertainty while the black dots are the points for this model that pass all constraints listed in Tab.~\ref{tab:obs}, other than $R_{K^{*0}}$.  This figure shows that our model cannot fit $R_{K^{*0}}$.  The calculated value of $R_{K^{*0}}$ from this model does not deviate significantly from the SM because the Wilson coefficients are multiplied by the SM-VL mixing squared, which is highly constrained by the electroweak precision measurements.

%%%%%%%%%%%%%%%%%%%%%%%%%%%%%%%%%%%%%%%%%%%%%%%%%%%%%%%%%%%%%%%%%%%%%%%%%%%%%
%%%%%%%%%%%%%%%%%%%%%%%%%%%%%%%%%%%%%%%%%%%%%%%%%%%%%%%%%%%%%%%%%%%%%%%%%%%%%
%%%%%%%%%%%%%%%%%%%%%%%%%%%%%%%%%%%%%%%%%%%%%%%%%%%%%%%%%%%%%%%%%%%%%%%%%%%%%

\section{Conclusions}

In this paper, we considered a very simple extension of the SM in which the SM is extended with one family of VL leptons, where the VL leptons couple to all three families of SM leptons.  We studied the constraints on this model coming from the heavy charged lepton mass bound, electroweak precision data, the muon $g-2$, lepton flavor violation, Higgs decays and lepton non-universality observables, $R_{^{*0}}$.  See Tab.~\ref{tab:obs} for the complete list of observables considered in this paper.  All experimental values, other than $R_{K^{*0}}$, are taken from Ref.~\cite{Olive:2016xmw}.  $R_{K^{*0}}$ is recently measured by LHCb~\cite{Bifani:2017lhcb}.  All theoretical calculations other than $R_{K^{*0}}$ are performed at leading order.  The NP Wilson coefficients contributing to $R_{K^{*0}}$ are computed at leading order while $R_{K^{*0}}$ is calculated using \texttt{flavio}~\cite{david_straub_2017_569011}.

In this paper, we showed that our model can fit all but the lepton non-universality measurement.  The most constraining observables are the muon $g-2, \ R_{\mu\mu}, \ R_{\gamma\gamma}$ and $\text{BR}(\mu\to e\gamma)$.  We find that if $R_{\mu\mu}$ is measured to be SM-like, then our model cannot simultaneously fit both the muon $g-2$ within $1\sigma$ and $R_{\mu\mu}$ (see Fig.~\ref{fig:mg2_vs_Rmumu}).  In addition, we also find that the SU(2) doublet VL mass is required to be $M_L\lesssim250\,\text{GeV}$ or $M_L\gtrsim400\,\text{GeV}$ in order to fit the muon $g-2$ within $1\sigma$ (see Fig.~\ref{fig:mg2_vs_ML}).  If in the future, the heavy charged lepton mass bound increases to be above $M_L\gtrsim250\,\text{GeV}$, then the muon $g-2$ can produce a stronger mass bound.  Fitting to the muon $g-2$ requires $||\lambda_\mu||>0.03$ while fitting to $R_{\gamma\gamma}$ requires $||\lambda_\mu||<0.09$.  Hence, the muon-VL coupling is constrained to be within $0.03<||\lambda_\mu||<0.09$.  Although we allow the VL leptons to couple to all three families of the SM leptons, by simultaneously fitting the muon $g-2$ and $\text{BR}(\mu\to e\gamma)$, the ratio of the electron-VL coupling to muon-VL coupling is constrained to be $\langle\lambda_e/\lambda_\mu\rangle\lesssim10^{-4}$.  Hence, this model requires some level of fine-tuning.  On the other hand, the strongest constraints on the tau-VL coupling is coming from electroweak precision observables.  The recently measured $\text{BR}(h\to\mu\tau)$ is less constraining than the electroweak precision observables.  We also find that this model cannot explain the lepton non-universality measurement.

%%%%%%%%%%%%%%%%%%%%%%%%%%%%%%%%%%%%%%%%%%%%%%%%%%%%%%%%%%%%%%%%%%%%%%%%%%%%%
%%%%%%%%%%%%%%%%%%%%%%%%%%%%%%%%%%%%%%%%%%%%%%%%%%%%%%%%%%%%%%%%%%%%%%%%%%%%%
%%%%%%%%%%%%%%%%%%%%%%%%%%%%%%%%%%%%%%%%%%%%%%%%%%%%%%%%%%%%%%%%%%%%%%%%%%%%%

\section*{Acknowledgments}
Z.P.~and S.R.~received partial support for this work from DOE/DE-SC0011726.  We would like to thank Andrzej Buras and Hong Zhang for discussions.

%%%%%%%%%%%%%%%%%%%%%%%%%%%%%%%%%%%%%%%%%%%%%%%%%%%%%%%%%%%%%%%%%%%%%%%%%%%%%
%%%%%%%%%%%%%%%%%%%%%%%%%%%%%%%%%%%%%%%%%%%%%%%%%%%%%%%%%%%%%%%%%%%%%%%%%%%%%
%%%%%%%%%%%%%%%%%%%%%%%%%%%%%%%%%%%%%%%%%%%%%%%%%%%%%%%%%%%%%%%%%%%%%%%%%%%%%
\appendix
\newpage
\section{Box Diagram Calculation}

In this appendix, we calculate the four box diagrams that have NP contributions to Wilson coefficient $C_9$ and $C_{10}$.  They are shown in Fig.~\ref{fig:box}.  To see all the Feynman rules explicitly, we start by rewriting part of the Lagrangian that is relevant to our calculation.  From Eq.~\ref{eq:W}, we have
\begin{align}
  \mathcal L\supset
  \frac{g}{\sqrt 2}\left[W_\mu^+\bar \nu_a\gamma^\mu([\tilde U_L]_{ab}P_L+[\tilde U_R]_{ab}P_R)\hat e_b
  + W_\mu^-\bar{\hat e}_b\gamma^\mu([\tilde U_L^*]_{ab}P_L+[\tilde U_R^*]_{ab}P_R)\nu_a\right] \,,
\end{align}
where $P_{L,R}$ are projection operators and
\begin{align}
  \tilde U_L &= \text{diag}(1,1,1,1,0)U_L \\
  \tilde U_R &= \text{diag}(0,0,0,1,0)U_R \,.
\end{align}
Notice that $[U_L]_{4a}=[\tilde U_L]_{4a}$ and similarly for $U_R$.

The relevant Lagrangian involving the Higgses are
\begin{align}
\begin{aligned}
  \mathcal L\supset
  &-\left(v+\frac{h}{\sqrt2}\right)\begin{pmatrix} \bar e_{L_i} & \bar L_L^- & \bar E_L \end{pmatrix}
  \begin{pmatrix}
    y_{ii}^e & 0 & \lambda_i^E \\
    \lambda_i^L & 0 & \lambda \\
    0 & \bar\lambda & 0
  \end{pmatrix}
  \begin{pmatrix} e_{R_i} \\ L_R^- \\ E_R \end{pmatrix} \\
  &-
  \frac{i\phi^0}{\sqrt2}\begin{pmatrix} \bar e_{L_i} & \bar L_L^- & \bar E_L \end{pmatrix}
  \begin{pmatrix}
    y_{ii}^e & 0 & \lambda_i^E \\
    \lambda_i^L & 0 & \lambda \\
    0 & -\bar\lambda & 0
  \end{pmatrix}
  \begin{pmatrix} e_{R_i} \\ L_R^- \\ E_R \end{pmatrix} \\
  &-
  \phi^+
  \begin{pmatrix} \bar\nu_{L_i} & \bar L_L^0 & 0 \end{pmatrix}
  \begin{pmatrix}
    y_{ii}^e & 0 & \lambda_i^E \\
    \lambda_i^L & 0 & \lambda \\
    0 & 0 & 0
  \end{pmatrix}
  \begin{pmatrix} e_{R_i} \\ L_R^- \\ E_R \end{pmatrix} \\
  &-
  \phi^-
  \begin{pmatrix} \bar e_{L_i} & \bar L_L^- & \bar E_L \end{pmatrix}
  \begin{pmatrix}
    0 & 0 & 0 \\
    0 & 0 & 0 \\
    0 & \bar\lambda & 0
  \end{pmatrix}
  \begin{pmatrix} 0_i \\ L_R^0 \\ 0 \end{pmatrix} + \text{h.c.} \,.
\end{aligned}
\end{align}
In the charged lepton mass basis, we have
\begin{align}
\begin{aligned}
  \mathcal L\supset
  &-\left(v+\frac{h}{\sqrt2}\right)\begin{pmatrix} \bar{\hat e}_{L_i} & \bar{\hat e}_{L_4} & \bar{\hat e}_{L_5} \end{pmatrix}
  \hat Y^e
  \begin{pmatrix} \hat e_{R_i} \\ \hat e_{R_4} \\ \hat e_{R_5} \end{pmatrix} \\
  &-
  \frac{i\phi^0}{\sqrt2}\begin{pmatrix} \bar{\hat e}_{L_i} & \bar{\hat e}_{L_4} & \bar{\hat e}_{L_5} \end{pmatrix}
  U_L^\dagger
  \begin{pmatrix}
    y_{ii}^e & 0 & \lambda_i^E \\
    \lambda_i^L & 0 & \lambda \\
    0 & -\bar\lambda & 0
  \end{pmatrix}
  U_R
  \begin{pmatrix} \hat e_{R_i} \\ \hat e_{R_4} \\ \hat e_{R_5} \end{pmatrix} \\
  &-
  \phi^+
  \begin{pmatrix} \bar\nu_{L_i} & \bar L_L^0 & 0 \end{pmatrix}
  \begin{pmatrix}
    y_{ii}^e & 0 & \lambda_i^E \\
    \lambda_i^L & 0 & \lambda \\
    0 & 0 & 0
  \end{pmatrix}
  U_R
  \begin{pmatrix} \hat e_{R_i} \\ \hat e_{R_4} \\ \hat e_{R_5} \end{pmatrix} \\
  &-
  \phi^-
  \begin{pmatrix} \bar{\hat e}_{L_i} & \bar{\hat e}_{L_4} & \bar{\hat e}_{L_5} \end{pmatrix}
  U_L^\dagger
  \begin{pmatrix}
    0 & 0 & 0 \\
    0 & 0 & 0 \\
    0 & \bar\lambda & 0
  \end{pmatrix}
  \begin{pmatrix} 0_i \\ L_R^0 \\ 0 \end{pmatrix} + \text{h.c.} \,.
\end{aligned}
\end{align}
The last two terms can be rewritten as
\begin{align}
  \mathcal L \supset
  -\phi^+\bar\nu_{L_b}[Y^{\nu_L}U_R]_{ba}\hat e_{R_a}
  -\phi^-\bar{\hat e}_{L_a}[U_L^\dagger Y^{\nu_R\dagger}]_{ab}\nu_{R_b} + \text{h.c.} \,.
\end{align}
where
\begin{align}
  Y^{\nu_L} \equiv
  \begin{pmatrix}
    y_{ii}^e & 0 & \lambda_i^E \\
    \lambda_i^L & 0 & \lambda \\
    0 & 0 & 0
  \end{pmatrix}
  \hspace{2em}\text{and}\hspace{2em}
  Y^{\nu_R\dagger} \equiv
  \begin{pmatrix}
    0 & 0 & 0 \\
    0 & 0 & 0 \\
    0 & \bar\lambda & 0
  \end{pmatrix} \,.
\end{align}
So, the coupling in diagrams (b)-(d) in Fig.~\ref{fig:box} involving $\phi^+$ are $-i([Y^{\nu_L}U_R]_{4a}P_R+[Y^{\nu_R}U_L]_{4a}P_L)$ while that involving $\phi^-$ are $-i([Y^{\nu_L*}U_R^*]_{4a}P_L+[Y^{\nu_R*}U_L^*]_{4a}P_R)$.

Since all the calculations are performed in the charged lepton mass basis, to simplify notation, we will drop $\hat{~}$ in the rest of the section.

Before we start to evaluate the four diagrams in Fig.~\ref{fig:box}, let's consider two loop integrals that we will be using.  These loop integrals are performed easily with \texttt{Package-X} developed by Patel~\cite{Patel:2015tea}.  The calculation is done in the `t Hooft-Feynman gauge.
\begin{align}
  A_{\alpha\beta}(M_i,M_L)
  &\equiv \int\frac{d^4q}{(2\pi)^4}\frac{q_\alpha q_\beta}{(q^2-M_W^2)^2(q^2-M_i^2)(q^2-M_L^2)}
  = -\frac{i}{64\pi^2M_W^2}g_1(x_i,y)g_{\alpha\beta} \,,
  \label{eq:loopA} \\
  B(M_i,M_L)
  &\equiv \int\frac{d^4q}{(2\pi)^4}\frac{1}{(q^2-M_W^2)^2(q^2-M_i^2)(q^2-M_L^2)}
  = -\frac{i}{16\pi^2M_W^4}g_0(x_i,y) \,,
  \label{eq:loopB}
\end{align}
where $x_i=M_i^2/M_W^2, \ y=M_L^2/M_W^2$ and
\begin{align}
  g_1(x,y) &= \frac{1}{x-y}\left[\frac{x^2}{(x-1)^2}\log x-\frac{y^2}{(y-1)^2}\log y-\frac{1}{x-1}+\frac{1}{y-1}\right] \,, \\
  g_0(x,y) &= \frac{1}{x-y}\left[\frac{x}{(x-1)^2}\log x-\frac{y}{(y-1)^2}\log y-\frac{1}{x-1}+\frac{1}{y-1}\right] \,.
\end{align}

\newpage
Diagram (a) in Fig.~\ref{fig:box} gives
\begin{align*}
  \Box^{(a)} =&\left(\frac{g}{\sqrt 2}\right)^4\sum_{i=u,c,t}V_{ib}V_{is}^*\int\frac{d^4q}{(2\pi)^4}\left(\frac{-i}{q^2-M_W^2}\right)^2 \left[\bar s\gamma^\mu P_L\frac{i(\slashed q+M_i)}{q^2-M_i^2}\gamma^\nu P_Lb\right] \\
  &\left[\bar e_a([U_L^*]_{4a}\gamma_\nu P_L+[U_R^*]_{4a}\gamma_\nu P_R)\frac{i(\slashed q+M_L)}{q^2-M_L^2}([U_L]_{4a}\gamma_\mu P_L+[U_R]_{4a}\gamma_\mu P_R)e_a\right] \\
  =& \frac{g^4}{4}\sum_{i=u,c,t}V_{ib}V_{is}^*\int\frac{d^4q}{(2\pi)^4}\frac{q_\alpha q_\beta}{(q^2-M_W^2)^2(q^2-M_i^2)(q^2-M_L^2)} \\
  &[\bar s\gamma^\mu P_L(\gamma^\alpha+M_i)\gamma^\nu P_Lb]  \\
  &\left[\bar e_a([U_L^*]_{4a}\gamma_\nu P_L+[U_R^*]_{4a}\gamma_\nu P_R)(\gamma^\beta+M_L)([U_L]_{4a}\gamma_\mu P_L+[U_R]_{4a}\gamma_\mu P_R)e_a\right] \,.
\end{align*}
Using Eq.~\ref{eq:loopA},
\begin{align*}
   \Box^{(a)} =& \frac{g^4}{4}\sum_{i=u,c,t}V_{ib}V_{is}^*A_{\alpha\beta}(M_i,M_L)[\bar s\gamma^\mu P_L(\gamma^\alpha+M_i)\gamma^\nu P_Lb] \\
  &[\bar e_a([U_L^*]_{4a}\gamma_\nu P_L+[U_R^*]_{4a}\gamma_\nu P_R)(\gamma^\beta+M_L)([U_L]_{4a}\gamma_\mu P_L+[U_R]_{4a}\gamma_\mu P_R)e_a] \,.
\end{align*}
The last two square brackets can be rewritten as
\begin{align*}
  & [\bar s\gamma^\mu\gamma^\alpha\gamma^\nu P_Lb][\bar e_a\gamma_\nu\gamma^\beta\gamma_\mu(|[U_L]_{4a}|^2P_L+|[U_R]_{4a}|^2P_R)e_a] \,,
\end{align*}
where we have dropped terms linear in $q$.  Using $g_{\alpha\beta}$ from $A_{\alpha\beta}$, we have
\begin{align*}
  & [\bar s\gamma^\mu\gamma^\alpha\gamma^\nu P_Lb][\bar e_a\gamma_\nu\gamma_\alpha\gamma_\mu(|[U_L]_{4a}|^2P_L+|[U_R]_{4a}|^2P_R)e_a] \,.
\end{align*}
Using the following Dirac matrix identity,
\begin{align*}
  \gamma^\mu\gamma^\alpha\gamma^\nu
  = g^{\mu\alpha}\gamma^\nu+g^{\alpha\nu}\gamma^\mu-g^{\mu\nu}\gamma^\alpha-i\epsilon^{\beta\mu\alpha\nu}\gamma_\beta\gamma^5 \,,
\end{align*}
we can rewrite the above equation as
\begin{align*}
  & 4[\bar s\gamma^\mu P_Lb][\bar e_a\gamma_\mu(|[U_L]_{4a}|^2P_L+|[U_R]_{4a}|^2P_R)e_a] \,.
\end{align*}
Putting all these together, we have
\begin{align}
  \Box^{(a)} =& -i\frac{4G_F}{\sqrt2}\sum_{i=u,c,t}V_{ib}V_{is}^*\frac{e^2}{16\pi^2}\frac{1}{s^2_W}\frac{1}{2}g_1(x_i,y)[\bar s\gamma^\mu P_Lb][\bar e_a\gamma_\mu(|[U_L]_{4a}|^2P_L+|[U_R]_{4a}|^2P_R)e_a] \,.
\end{align}
Hence, the contribution of this diagram to the Wilson coefficients are
\begin{align}
  C_9^\text{NP(a)} &= -\frac{1}{s^2_W}\frac{1}{4}(|[U_L]_{4a}|^2+|[U_R]_{4a}|^2)g_1(x_i,y) \,, \\
  C_{10}^\text{NP(a)} &= \frac{1}{s^2_W}\frac{1}{4}(|[U_L]_{4a}|^2-|[U_R]_{4a}|^2)g_1(x_i,y) \,.
\end{align}

\newpage
Diagram (b) in Fig.~\ref{fig:box} gives
\begin{align*}
  \Box^{(b)} =& \left(\frac{g}{\sqrt 2}\right)^4\sum_{i=u,c,t}V_{ib}V_{is}^*\int\frac{d^4q}{(2\pi)^4}\left(\frac{-i}{q^2-M_W^2}\right)^2\left[\bar s\gamma^\mu P_L\frac{i(\slashed q+M_i)}{q^2-M_i^2}\frac{M_i}{M_W}P_Lb\right] \\
  &\left[\bar e_a\frac{-v([Y^{\nu_L*}U_R^*]_{4a}P_L+[Y^{\nu_R*}U_L^*]_{4a}P_R)}{M_W}\frac{i(\slashed q+M_L)}{q^2-M_L^2}([U_L]_{4a}\gamma_\mu P_L+[U_R]_{4a}\gamma_\nu P_R)e_a\right] \\
  =& \frac{g^4}{4}\sum_{i=u,c,t}V_{ib}V_{is}^*\int\frac{d^4q}{(2\pi)^4}\frac{1}{(q^2-M_W^2)^2(q^2-M_i^2)(q^2-M_L^2)} \\
  &\left[\bar s\gamma^\mu P_L(\slashed q+M_i)\frac{M_i}{M_W}P_Lb\right] \\
  &\left[\bar e_a\frac{-v([Y^{\nu_L*}U_R^*]_{4a}P_L+[Y^{\nu_R*}U_L^*]_{4a}P_R)}{M_W}(\slashed q+M_L)([U_L]_{4a}\gamma_\mu P_L+[U_R]_{4a}\gamma_\mu P_R)e_a\right] \,,
\end{align*}
where we have neglected external masses.  Using Eq.~\ref{eq:loopB},
\begin{align*}
  \Box^{(b)} =& \frac{g^4}{4}\sum_{i=u,c,t}V_{ib}V_{is}^*B(M_i,M_L)\left[\bar s\gamma^\mu P_L(\slashed q+M_i)\frac{M_i}{M_W}P_Lb\right] \\
  &\left[\bar e_a\frac{-v([Y^{\nu_L*}U_R^*]_{4a}P_L+[Y^{\nu_R*}U_L^*]_{4a}P_R)}{M_W}(\slashed q+M_L)([U_L]_{4a}\gamma_\mu P_L+[U_R]_{4a}\gamma_\mu P_R)e_a\right] \,,
\end{align*}
The last two square brackets can be rewritten as
\begin{align*}
  -\frac{vM_i^2M_L}{M_W^2}[\bar s\gamma^\mu P_Lb][\bar e_a\gamma_\mu([U_L]_{4a}[Y^{\nu_R*}U_L^*]_{4a}P_L+[U_R]_{4a}[Y^{\nu_L*}U_R^*]_{4a}P_R)e_a] \,,
\end{align*}
where we have dropped terms linear in $q$.  Putting all these together, we have
\begin{align}
\begin{aligned}
  \Box^{(b)} =& i\frac{4G_F}{\sqrt2}\sum_{i=u,c,t}V_{ib}V_{is}^*\frac{e^2}{16\pi^2}\frac{1}{s^2_W}\frac{1}{2}\frac{v}{M_L}x_iyg_0(x_i,y) \\
  &[\bar s\gamma^\mu P_Lb][\bar e_a\gamma_\mu([U_L]_{4a}[Y^{\nu_R*}U_L^*]_{4a}P_L+[U_R]_{4a}[Y^{\nu_L*}U_R^*]_{4a}P_R)e_a] \,.
\end{aligned}
\end{align}
Hence, the contribution of this diagram to the Wilson coefficients are
\begin{align}
  C_9^\text{NP(b)} &= \frac{1}{s^2_W}\frac{1}{4}\frac{v}{M_L}x_iy([U_L]_{4a}[Y^{\nu_R*}U_L^*]_{4a}+[U_R]_{4a}[Y^{\nu_L*}U_R^*]_{4a})g_0(x_i,y) \,,\\
  C_{10}^\text{NP(b)} &= -\frac{1}{s^2_W}\frac{1}{4}\frac{v}{M_L}x_iy([U_L]_{4a}[Y^{\nu_R*}U_L^*]_{4a}-[U_R]_{4a}[Y^{\nu_L*}U_R^*]_{4a})g_0(x_i,y) \,.
\end{align}

\newpage
Diagram (c) in Fig.~\ref{fig:box} gives
\begin{align*}
  \Box^{(c)} =& \left(\frac{g}{\sqrt 2}\right)^4\sum_{i=u,c,t}V_{ib}V_{is}^*\int\frac{d^4q}{(2\pi)^4}\left(\frac{-i}{p^2-M_W^2}\right)^2 \left[\bar s\frac{M_i}{M_W}P_R\frac{i(\slashed q+M_i)}{q^2-M_i^2}\gamma^\mu P_Lb\right] \\
  &\left[\bar e_a([U_L^*]_{4a}\gamma_\mu P_L+[U_R^*]_{4a}\gamma_\mu P_R)\frac{i(\slashed q+M_L)}{q^2-M_L^2}\frac{-v([Y^{\nu_L}U_R]_{4a}P_R+[Y^{\nu_R}U_L]_{4a}P_L)}{M_W}e_a\right] \\
  =& \frac{g^2}{4}\sum_{i=u,c,t}V_{ib}V_{is}^*\int\frac{d^4q}{(2\pi)^4}\frac{1}{(q^2-M_W^2)^2(q^2-M_i^2)(q^2-M_L^2)}  \\
  & \left[\bar s\frac{M_i}{M_W}P_R(\slashed q+M_i)\gamma^\mu P_Lb\right] \\
  &\left[\bar e_a([U_L^*]_{4a}\gamma_\mu P_L+[U_R^*]_{4a}\gamma_\mu P_R)(\slashed q+M_L)\frac{-v([Y^{\nu_L}U_R]_{4a}P_R+[Y^{\nu_R}U_L]_{4a}P_L)}{M_W}e_a\right] \,,
\end{align*}
where we have neglected external masses.  Using Eq.~\ref{eq:loopB},
\begin{align*}
  \Box^{(c)} =& \frac{g^4}{4}\sum_{i=u,c,t}V_{ib}V_{is}^*B(M_i,M_L)\left[\bar s\frac{M_i}{M_W}P_R(\slashed q+M_i)\gamma^\mu P_Lb\right] \\
  &\left[\bar e_a([U_L^*]_{4a}\gamma_\mu P_L+[U_R^*]_{4a}\gamma_\mu P_R)(\slashed q+M_L)\frac{-v([Y^{\nu_L}U_R]_{4a}P_R+[Y^{\nu_R}U_L]_{4a}P_L)}{M_W}e_a\right] \,.
\end{align*}
The last two square brackets can be rewritten as
\begin{align*}
  -\frac{vM_i^2M_L}{M_W^2}[\bar s\gamma^\mu P_Lb][\bar e_a\gamma_\mu([U_L^*]_{4a}[Y^{\nu_R}U_L]_{4a}P_L+[U_R^*]_{4a}[Y^{\nu_L}U_R]_{4a}P_R)e_a] \,,
\end{align*}
where we have dropped terms linear in $q$.  Putting all these together, we have
\begin{align}
\begin{aligned}
  \Box^{(c)} =& i\frac{4G_F}{\sqrt2}\sum_{i=u,c,t}V_{ib}V_{is}^*\frac{e^2}{16\pi^2}\frac{1}{s^2_W}\frac{1}{2}\frac{v}{M_L}x_iyg_0(x_i,y) \\
  &[\bar s\gamma^\mu P_Lb][\bar e_a\gamma_\mu([U_L^*]_{4a}[Y^{\nu_R}U_L]_{4a}P_L+[U_R^*]_{4a}[Y^{\nu_L}U_R]_{4a}P_R)e_a] \,.
\end{aligned}
\end{align}
Hence, the contribution of this diagram to the Wilson coefficients are
\begin{align}
  C_9^\text{NP(c)} &= \frac{1}{s^2_W}\frac{1}{4}\frac{v}{M_L}x_iy([U_L^*]_{4a}[Y^{\nu_R}U_L]_{4a}+[U_R^*]_{4a}[Y^{\nu_L}U_R]_{4a})g_0(x_i,y) \,,\\
  C_{10}^\text{NP(c)} &= -\frac{1}{s^2_W}\frac{1}{4}\frac{v}{M_L}x_iy([U_L^*]_{4a}[Y^{\nu_R}U_L]_{4a}-[U_R^*]_{4a}[Y^{\nu_L}U_R]_{4a})g_0(x_i,y) \,.
\end{align}

\newpage
Diagram (d) in Fig.~\ref{fig:box} gives
\begin{align*}
  \Box^{(d)} =& \left(\frac{g}{\sqrt 2}\right)^4\sum_{i=u,c,t}V_{ib}V_{is}^*\int\frac{d^4q}{(2\pi)^4}\left(\frac{-i}{q^2-M_W^2}\right)^2 \left[\bar s\frac{M_i}{M_W}P_R\frac{i(\slashed q+M_i)}{q^2-M_i^2}\frac{M_i}{M_W}P_Lb\right] \\
  &\left[\bar e_a\frac{v([Y^{\nu_L*}U_R^*]_{4a}P_L+[Y^{\nu_R*}U_L^*]_{4a}P_R)}{M_W}\frac{i(\slashed q+M_L)}{q^2-M_L^2}\frac{v([Y^{\nu_L}U_R]_{4a}P_R+[Y^{\nu_R}U_L]_{4a}P_L)}{M_W}e_a\right] \\
  =& \frac{g^4}{4}\sum_{i=u,c,t}V_{ib}V_{is}^*\int\frac{d^4q}{(2\pi)^4}\frac{q_\alpha q_\beta}{(q^2-M_W^2)(q^2-M_i^2)(q^2-M_L^2)} \\
  & \left[\bar s\frac{M_i}{M_W}P_R(\gamma^\alpha+M_i)\frac{M_i}{M_W}P_Lb\right] \\
  &\left[\bar e_a\frac{v([Y^{\nu_L*}U_R^*]_{4a}P_L+[Y^{\nu_R*}U_L^*]_{4a}P_R)}{M_W}(\gamma^\beta+M_L)\frac{v([Y^{\nu_L}U_R]_{4a}P_R+[Y^{\nu_R}U_L]_{4a}P_L)}{M_W}e_a\right] \,,
\end{align*}
where we have neglected external masses.  Using Eq.~\ref{eq:loopA},
\begin{align*}
  \Box^{(d)} =& \frac{g^4}{4}\sum_{i=u,c,t}V_{ib}V_{is}^*A_{\alpha\beta}(M_i,M_L)\left[\bar s\frac{M_i}{M_W}P_R(\gamma^\alpha+M_i)\frac{M_i}{M_W}P_Lb\right] \\
  &\left[\bar e_a\frac{v([Y^{\nu_L*}U_R^*]_{4a}P_L+[Y^{\nu_R*}U_L^*]_{4a}P_R)}{M_W}(\gamma^\beta+M_L)\frac{v([Y^{\nu_L}U_R]_{4a}P_R+[Y^{\nu_R}U_L]_{4a}P_L)}{M_W}e_a\right] \,,
\end{align*}
Using $g_{\alpha\beta}$ from $A_{\alpha\beta}$, the last two square brackets can be rewritten as
\begin{align*}
  \frac{v^2M_i^2}{M_W^4}[\bar s\gamma^\mu P_Lb][\bar e_a\gamma_\mu(|[Y^{\nu_R}U_L]_{4a}|^2P_L+|[Y^{\nu_L}U_R]_{4a}|^2P_R)e_a] \,,
\end{align*}
where we have dropped terms linear in $q$.  Putting all these together, we have
\begin{align}
\begin{aligned}
  \Box^{(d)} =& -i\frac{4G_F}{\sqrt2}\sum_{i=u,c,t}V_{ib}V_{is}^*\frac{e^2}{16\pi^2}\frac{1}{s^2_W}\frac{1}{8}\frac{v^2}{M_L^2}x_iyg_1(x_i,y) \\
  & [\bar s\gamma^\mu P_Lb] [\bar e_a\gamma_\mu(|[Y^{\nu_R}U_L]_{4a}|^2P_L+|[Y^{\nu_L}U_R]_{4a}|^2P_R)e_a] \,.
\end{aligned}
\end{align}
Hence, the contribution of this diagram to the Wilson coefficients are
\begin{align}
  C_9^\text{NP(d)} &= -\frac{1}{s^2_W}\frac{1}{16}\frac{v^2}{M_L^2}x_iy(|[Y^{\nu_R}U_L]_{4a}|^2+|[Y^{\nu_L}U_R]_{4a}|^2)g_1(x_i,y) \,,\\
  C_{10}^\text{NP(d)} &= \frac{1}{s^2_W}\frac{1}{16}\frac{v^2}{M_L^2}x_iy(|[Y^{\nu_R}U_L]_{4a}|^2-|[Y^{\nu_L}U_R]_{4a}|^2)g_1(x_i,y) \,.
\end{align}

\newpage
The total contribution to the Wilson coefficients is the sum of the contribution from the four diagrams:
\begin{align}
\begin{aligned}
  C_9^\text{NP} =& -\frac{1}{s^2_W}\frac{1}{4}\bigg[\left(|[U_L]_{4a}|^2+|[U_R]_{4a}|^2+\frac{1}{4}\frac{v^2}{M_L^2}x_iy(|[Y^{\nu_R}U_L]_{4a}|^2+|[Y^{\nu_L}U_R]_{4a}|^2)\right)g_1(x_i,y) \\
  &\hspace{3em}-\frac{v}{M_L}x_iy([U_L]_{4a}[Y^{\nu_R*}U_L^*]_{4a}+[U_L^*]_{4a}[Y^{\nu_R}U_L]_{4a} \\
  &\hspace{9em}+[U_R]_{4a}[Y^{\nu_L*}U_R^*]_{4a}+[U_R^*]_{4a}[Y^{\nu_L}U_R]_{4a})g_0(x_i,y)\bigg] \,,
\end{aligned} \\
\begin{aligned}
  C_{10}^\text{NP} =& \frac{1}{s^2_W}\frac{1}{4}\bigg[\left(|[U_L]_{4a}|^2-|[U_R]_{4a}|^2+\frac{1}{4}\frac{v^2}{M_L^2}x_iy(|[Y^{\nu_R}U_L]_{4a}|^2-|[Y^{\nu_L}U_R]_{4a}|^2)\right)g_1(x_i,y) \\
  &\hspace{3em}-\frac{v}{M_L}x_iy([U_L]_{4a}[Y^{\nu_R*}U_L^*]_{4a}+[U_L^*]_{4a}[Y^{\nu_R}U_L]_{4a} \\
  &\hspace{9em}-[U_R]_{4a}[Y^{\nu_L*}U_R^*]_{4a}-[U_R^*]_{4a}[Y^{\nu_L}U_R]_{4a})g_0(x_i,y)\bigg] \,.
\end{aligned}
\end{align}

%%%%%%%%%%%%%%%%%%%%%%%%%%%%%%%%%%%%%%%%%%%%%%%%%%%%%%%%%%%%%%%%%%%%%%%%%%%%%
%%%%%%%%%%%%%%%%%%%%%%%%%%%%%%%%%%%%%%%%%%%%%%%%%%%%%%%%%%%%%%%%%%%%%%%%%%%%%
%%%%%%%%%%%%%%%%%%%%%%%%%%%%%%%%%%%%%%%%%%%%%%%%%%%%%%%%%%%%%%%%%%%%%%%%%%%%%
\clearpage
\newpage
\bibliographystyle{utphys}
\bibliography{bibliography}

\end{document}